\documentclass[11pt]{article}
\usepackage{amssymb,amsmath,amsfonts}
\usepackage{graphicx}
\usepackage{graphics}
\usepackage{eepic,epsfig}

\begin{document}

\title{Electromagnetic two-point functions and the Casimir effect \\
in Friedmann-Robertson-Walker cosmologies}
\author{ S. Bellucci$^{1}$\thanks{%
E-mail: bellucci@lnf.infn.it }, A. A. Saharian$^{2}$\thanks{%
E-mail: saharian@ysu.am }\vspace{0.3cm} \\
%EndAName
\textit{$^1$ INFN, Laboratori Nazionali di Frascati,}\\
\textit{Via Enrico Fermi 40, 00044 Frascati, Italy} \vspace{0.3cm}\\
\textit{$^2$ Department of Physics, Yerevan State University,}\\
\textit{1 Alex Manoogian Street, 0025 Yerevan, Armenia }}
\maketitle

\begin{abstract}
We evaluate the two-point functions of the electromagnetic field in $(D+1)$%
-dimensional spatially flat Friedmann-Robertson-Walker universes with a
power-law scale factor, assuming that the field is prepared in the
Bunch-Davies vacuum state. The range of powers are specified in which the
two-point functions are infrared convergent and the Bunch-Davies vacuum for
the electromagnetic field is a physically realizable state. The two-point
functions are applied for the investigation of the vacuum expectation values
of the field squared and the energy-momentum tensor, induced by a single and
two parallel conducting plates. Unlike to the case of conducting plates in
the Minkowski bulk, in the problem under consideration the stresses along
the directions parallel to the plates are not equal to the energy density.
We show that, in addition to the diagonal components, the vacuum
energy-momentum tensor has a nonzero off-diagonal component which describes
energy flux along the direction normal to the plates. For a single plate
this flux is directed from the plate. The Casimir forces are investigated in
the geometry of two plates. At separations between the plates smaller than
the curvature radius of the background spacetime, to the leading order, we
recover the corresponding result in the Minkowski spacetime and in this case
the forces are attractive. At larger separations, the influence of the
curvature on the Casimir forces is essential with different asymptotic
behavior for decelerated and accelerated expansions. In particular, for the
latter case there is a range of powers of the expansion law in which the
forces become repulsive at large separations between the plates.
\end{abstract}

\bigskip


\bigskip

\section{Introduction}

Observational data on the large-scale structure of the Universe and on the
cosmic microwave background indicate that at large scales the Universe is
homogeneous and isotropic and its geometry is well described by the
Friedmann-Robertson-Walker (FRW) metric. Because of the high symmetry,
numerous physical problems are exactly solvable on this background and a
better understanding of physical effects in FRW models could serve as a
handle to deal with more complicated geometries. In particular, the
investigation of quantum effects at the early stages of the cosmological
expansion has been a subject of study in many research papers (see \cite%
{Birr82} for early investigations and \cite{Boya94}-\cite{Beze11} and
references therein for later developments). Physical effects such as the
polarization of the vacuum and particle creation by strong gravitational
fields can have a profound impact on the dynamics of the expansion. They
provide a natural mechanism for the damping of anisotropies in the early
Universe and for a solution of the cosmological singularity problem. During
an inflationary epoch, quantum fluctuations introduce inhomogeneities which
explain the origin of the primordial density perturbations needed to explain
the formation of the large-scale structure in the Universe \cite{Lind90}.

The nontrivial properties of the vacuum are among the most interesting
predictions of the quantum field theory. They are sensitive to both the
local and global characteristics of the background spacetime in the early
Universe. In particular, the properties of the vacuum state can be changed
by the imposition of boundary conditions on the field operator. These
conditions may be induced either by the presence of physical boundaries,
like material boundaries in QED, or by the nontrivial topology of the
background space.\ The both types of boundary conditions give rise to the
modification of the spectrum for the vacuum fluctuations of a quantum field
and, as a result, to the change in the vacuum expectation values (VEVs) of
various physical observables. This effect for the electromagnetic field was
first predicted by Casimir in 1948 and recently it has been experimentally
confirmed with great precision (for reviews of the Casimir effect see Ref.
\cite{Eliz94}).

The explicit dependence of the VEVs on the bulk and boundary geometries can
be found for highly symmetric backgrounds only. In particular, motivated by
Randall-Sundrum-type braneworld models, investigations of the Casimir effect
in anti-de Sitter spacetime have attracted a great deal of attention. In
these models, the presence of the branes imposes boundary conditions on bulk
quantum fields and, as a consequence of this, the Casimir-type forces arise
acting on the branes. This provides a natural mechanism for the
stabilization of the interbrane distance (radion), as required for a
solution of the hierarchy problem. In addition, the Casimir energy
contributes to both the brane and bulk cosmological constants and it has to
be taken into account in any self-consistent formulation of the braneworld
dynamics. The Casimir energy and forces in the geometry \ of two parallel
branes in anti-de Sitter bulk have been investigated in Refs.~\cite{Gold00}
by using either dimensional or zeta function regularization methods. Local
Casimir densities for scalar and fermionic fields were studied in Ref.~\cite%
{Knap04}. Higher-dimensional braneworld models in anti-de Sitter spacetime
with compact internal spaces are considered in~Ref. \cite{Flac03}. Another
maximally symmetric gravitational background which plays an important role
in cosmology is de Sitter spacetime. For a massive scalar field with general
curvature coupling parameter, the Casimir densities in this background,
induced by flat and spherical boundaries with Robin boundary conditions,
have been investigated in Ref. \cite{Saha09} (for special cases of
conformally and minimally coupled massless fields see Ref. \cite{Seta01}).
More recently, the electromagnetic vacuum densities induced by a conducting
plate in de Sitter spacetime have been discussed in Ref. \cite{Saha13}. In
both cases of anti-de Sitter and de Sitter bulks, the curvature of the
background spacetime decisively influences the properties of the vacuum at
distances from the boundaries larger than the curvature scale of the
background spacetime. Similar features were observed in the topological
Casimir effect for scalar and fermionic fields, induced by toroidal
compactification of spatial dimensions in de Sitter spacetime \cite{Saha08}.

In the present paper we consider an exactly solvable problem for combined
effects of the gravitational field and boundaries on the properties of the
electromagnetic vacuum in a more general cosmological background. The
two-point functions, the expectation values of the field squared and the
energy-momentum tensor in the Bunch-Davies vacuum state are evaluated for
the geometry of two parallel conducting plates in spatially flat FRW
universes where the scale factor is a power of the comoving time (for the
vacuum polarization and the particle creation in FRW cosmological models
with power law scale factors see Refs. \cite{Bord97,Bunc78,Ford77}). The
vacuum expectation values of the energy-momentum tensor and the Casimir
forces for conformally invariant scalar and electromagnetic fields in the
geometry of curved boundaries on the background of FRW spacetime with
negative spatial curvature have been recently investigated in Ref. \cite%
{Saha10}. In these papers, the boundaries are conformal images of flat
boundaries in Rindler spacetime and the conformal relation between the FRW
and Rindler spacetimes was used. Taking into account the recent interest to
higher-dimensional models of both Kaluza-Klein and braneworld types and
prospective applications in multidimensional cosmology, in the present paper
we consider the case of arbitrary number of spatial dimensions $D$. The
electromagnetic field is conformally invariant in $D=3$ and in this case the
Casimir densities and forces for the Bunch-Davies vacuum state are simply
obtained from the corresponding results in the Minkowski bulk by conformal
transformation. In other dimensions this is not the case and the curvature
of background spacetime leads to new qualitative features in the behavior of
the vacuum characteristics. In particular, as it will be shown below, a
nonzero energy flux appears along the direction normal to the plates.

The structure the paper is the following. In the next section we describe
the background geometry and present a complete set of mode functions for the
vector potential of the electromagnetic field. Then, these mode functions
are used for the evaluation of the two-point functions for the vector
potential and field tensor. In section \ref{sec:TPpl}, the mode functions
and the two-point functions are derived in the region between two conducting
plates. Section \ref{sec:Cas1pl} is devoted to the investigation of the VEVs
for the field squared and the energy-momentum tensor in the geometry of a
single plate. The asymptotic behavior of the VEVs at small and large
distances from the plate is discussed in detail. The VEVs in the geometry of
two plates and the Casimir interaction forces are investigated in section %
\ref{sec:2plates}. The main results of the paper are summarized in section %
\ref{sec:Conc}.

\section{Electromagnetic modes and two-point functions in boundary-free FRW
spacetime}

\label{sec:TP}

\subsection{Background geometry}

As background geometry we consider a spatially flat $(D+1)$-dimensional FRW
spacetime with a power law scale factor versus time:%
\begin{equation}
ds^{2}=dt^{2}-a^{2}(t)(d\mathbf{z})^{2},\;a(t)=a_{1}t^{c},  \label{ds2}
\end{equation}%
where $\mathbf{z}=(z^{1},\ldots ,z^{D})$ are comoving spatial coordinates
and $0<t<\infty $. For the scalar curvature corresponding to the line
element (\ref{ds2}) one has $R=Dc\left[ (D+1)c-2\right] /t^{2}$. Below, in
addition to the comoving time coordinate $t$, it is convenient to introduce
the conformal time defined as $\tau =\int dt/a(t)$. In terms of the
conformal time, for $c\neq 1$, one has%
\begin{equation}
a(\tau )=a_{0}\eta ^{b},\;t=\left[ a_{1}(1-c)\tau \right] ^{1/(1-c)},
\label{taut}
\end{equation}%
where we have introduced the notations%
\begin{equation}
\eta =|\tau |,\;b=c/(1-c),\;a_{0}=a_{1}(a_{1}|1-c|)^{b}.  \label{b}
\end{equation}%
Note that for $0<c<1$ one has $0<\tau <\infty $, and for $c>1$ one has $%
-\infty <\tau <0$. With this time coordinate, the line element is written in
a conformally flat form: $ds^{2}=a^{2}(\tau )[d\tau ^{2}-(d\mathbf{z})^{2}]$%
. Taking $c=0$ and $a_{1}=1$, from the formulas given below we obtain the
corresponding results in the Minkowski bulk. In another limiting case,
corresponding to $c\rightarrow \infty $, one has $b=-1$ and the scale factor
(\ref{taut}) reduces to the one for de Sitter spacetime: $a(\tau
)=a_{0}/\eta $ (in comoving time $a(t)=e^{t/a_{0}}$). In the case $c=1$ for
the conformal time we have $\tau =a_{1}^{-1}\ln (t/t_{0})$ with the scale
factor $a(\tau )=a_{1}t_{0}e^{a_{1}\tau }$, $-\infty <\tau <+\infty $. In
the discussion below we shall assume that $c\neq 1$.

The energy density $\rho $ and the pressure $p$ corresponding to the line
element (\ref{ds2}) are found from the Einstein equations:%
\begin{equation}
\rho =(D-1)\frac{Dc^{2}}{16\pi Gt^{2}},\;p=\left( D-1\right) \frac{c(2-cD)}{%
16\pi Gt^{2}},  \label{rope}
\end{equation}%
with $G$ being the $(D+1)$-dimensional gravitational constant. Note that the
energy density is always nonnegative, whereas the pressure can be either
positive or negative. The equation of state for the source (\ref{rope}) is
of barotropic type, $p=w\rho $, with the equation of state parameter $%
w=2/(cD)-1$. In the special case $c=2/(D+1)$ the Ricci scalar vanishes and
this corresponds to a source of the radiation type. For $c=2/D$ the pressure
is zero and one has another important special case of dust-matter driven
models. In the case $c>1$, the expansion described by the scale factor (\ref%
{ds2}) is accelerating. This type of expansion is employed in\ power law
\cite{Abbo87} and extended \cite{La89} inflationary models. The
corresponding dynamics may be driven by a scalar field $\varphi $ with an
exponential potential $V(\varphi )\propto e^{-\beta \varphi }$. In this case
the parameter $c$ is given by the expression $c=4\beta ^{-2}/(D-1)$ (see,
for instance, \cite{Burd88}). Power law solutions with $c=2(n-1)(2n-1)/(D-2n)
$ arise in higher-order gravity theories with the Lagrangian $R^{n}$ \cite%
{Maed88} . The results given below can also be applied to cosmological
models driven by phantom energy for which $w<-1$. For the corresponding
scale factor one has $a(t)=a_{1}(-t)^{c}$, $-\infty <t<0$, with $c<0$ and
with the relation $t=-\left[ a_{1}(c-1)\tau \right] ^{1/(1-c)}$, $-\infty
<\tau <0$. Note that we have a decelerating expansion for $0<c<1$ and an
accelerating expansion for $c<0$ or $c>1$.

\subsection{Mode functions}

We consider the electromagnetic field with the action integral%
\begin{equation}
S=-\frac{1}{16\pi }\int d^{D+1}x\,\sqrt{|g|}F_{\mu \nu }(x)F^{\mu \nu }(x),
\label{action}
\end{equation}%
in background described by the line element (\ref{ds2}). In Eq. (\ref{action}%
), $F_{\mu \nu }=\partial _{\mu }A_{\nu }-\partial _{\nu }A_{\mu }$ is the
electromagnetic field tensor and $g$ is the determinant of the metric
tensor. For the vector potential $A_{\mu }$ the gauge conditions $A_{0}=0$,$%
\;\nabla _{\mu }A^{\mu }=0$ will be imposed. The background geometry is
spatially flat and we expand the vector potential in Fourier integral%
\begin{equation}
A_{l}(x)=\int d\mathbf{k}\,A_{l}(\tau ,\mathbf{k})e^{i\mathbf{k}\cdot
\mathbf{z}},\;A_{l}(\tau ,-\mathbf{k})=A_{l}^{\ast }(\tau ,\mathbf{k}),
\label{AFour}
\end{equation}%
with shorthand notations $x=(\tau ,\mathbf{z})$, $\mathbf{k}=(k_{1},\ldots
,k_{D})$ and $\mathbf{k}\cdot \mathbf{z}=\sum_{l=1}^{D}k_{l}z^{l}$. From the
gauge condition one has the relation $\sum_{l=1}^{D}k_{l}A_{l}(\tau ,\mathbf{%
k})=0$.

Substituting the expansion (\ref{AFour}) into the action integral and
integrating over the spatial coordinates, one gets%
\begin{equation}
S=\frac{(2\pi )^{D-1}}{4}\sum_{l=1}^{D}\int d\mathbf{k}\,\int d\tau
\,a^{D-3}(\tau )\left[ |\partial _{\tau }A_{l}(\tau ,\mathbf{k}%
)|^{2}-k^{2}|A_{l}(\tau ,\mathbf{k})|^{2}\right] ,  \label{Action2}
\end{equation}%
with $k=|\mathbf{k}|$. This action gives rise to the following equation for
the Fourier modes of the vector potential:%
\begin{equation}
\partial _{\tau }\left[ a^{D-3}(\tau )\partial _{\tau }A_{l}(\tau ,\mathbf{k}%
)\right] +a^{D-3}(\tau )k^{2}A_{l}(\tau ,\mathbf{k})=0.  \label{EqA}
\end{equation}%
For the scale factor given by Eq. (\ref{taut}), the general solution of Eq. (%
\ref{EqA}) is a linear combination of $\eta ^{\nu }H_{\nu }^{(1)}(k\tau )$
and $\eta ^{\nu }H_{\nu }^{(2)}(k\tau )$, where $H_{\nu }^{(1,2)}(x)$ are
the Hankel functions and%
\begin{equation}
\nu =\frac{1}{2}\left[ 1+\frac{(D-3)c}{c-1}\right] .  \label{nu}
\end{equation}%
One of the coefficients in the linear combination is determined from the
normalization condition for the mode functions and the second coefficient is
fixed by the choice of the vacuum state. In what follows we shall assume
that the electromagnetic field is prepared in the Bunch-Davies vacuum state
\cite{Bunc78} for which $A_{l}(\tau ,\mathbf{k})\sim \eta ^{\nu }H_{\nu
}^{(2)}(k\tau )$. With this choice, the two-point functions admit the
Hadamard form. In the limit of slow expansion the Bunch-Davies vacuum is
reduced to the standard Minkowskian vacuum. In the discussion below, we
shall write the mode functions in terms of the Macdonald function by using
the relation $H_{\nu }^{(2)}(k\tau )=(2i/\pi )e^{\nu \pi i/2}K_{\nu }(k\eta
e^{s_{\tau }\pi i/2})$, where $s_{\tau }=\mathrm{sgn}(\tau )$. Note that $%
s_{\tau }=1$ for $0<c\,<1$ and $s_{\tau }=-1$ for other cases.

Hence, in the quantization procedure of the electromagnetic field as a
complete set of mode functions we can take
\begin{equation}
A_{(\sigma \mathbf{k})l}(x)=C_{0}\epsilon _{(\sigma )l}\eta ^{\nu }K_{\nu
}(k\eta e^{s_{\tau }\pi i/2})e^{i\mathbf{k}\cdot \mathbf{z}},\text{ }%
l=1,\ldots ,D,  \label{Amodes}
\end{equation}%
with the polarization vectors $\epsilon _{(\sigma )l}$, $\sigma =1,\ldots
,D-1$. For these vectors one has the relations%
\begin{equation}
\sum_{l=1}^{D}\epsilon _{(\sigma )l}k_{l}=0,\;\sum_{l=1}^{D}\epsilon
_{(\sigma )l}\epsilon _{(\sigma ^{\prime })l}=\delta _{\sigma \sigma
^{\prime }},  \label{Eps1}
\end{equation}%
and%
\begin{equation}
\sum_{\sigma =1}^{D-1}\epsilon _{(\sigma )l}\epsilon _{(\sigma )m}=\delta
_{lm}-k_{l}k_{m}/k^{2}.  \label{Eps2}
\end{equation}%
The coefficient $C_{0}$ in Eq. (\ref{Amodes}) is determined from the
normalization condition:%
\begin{equation}
\ \int d\mathbf{z}\,\sum_{l=1}^{D}[A_{(\sigma \mathbf{k})l}^{\ast
}(x)\partial _{\tau }\ A_{(\sigma ^{\prime }\mathbf{k}^{\prime
})l}(x)-A_{(\sigma ^{\prime }\mathbf{k}^{\prime })_{l}}(x)\partial _{\tau }\
A_{(\sigma \mathbf{k})l}^{\ast }(x)]=\frac{4\pi \delta _{\sigma \sigma
^{\prime }}}{ia^{D-3}}\delta (\mathbf{k}-\mathbf{k}^{\prime }).
\label{Ncond}
\end{equation}%
By making use of the relation%
\begin{equation}
K_{\nu }(k\eta e^{-s_{\tau }\pi i/2})=e^{s_{\tau }\nu \pi i}K_{\nu }(k\eta
e^{s_{\tau }\pi i/2})+s_{\tau }\pi iI_{\nu }(k\eta e^{s_{\tau }\pi i/2}),
\label{rel1}
\end{equation}%
with $I_{\nu }(x)$ being the modified Bessel function, the time-dependent
part in the normalization integral is reduced to the Wronskian $W\{K_{\nu
}(k\eta e^{s_{\tau }\pi i/2}),I_{\nu }(k\eta e^{s_{\tau }\pi i/2})\}$. In
this way, for the normalization factor one gets%
\begin{equation}
\left\vert C_{0}\right\vert ^{2}=\frac{4a_{0}^{3-D}}{\left( 2\pi \right) ^{D}%
}.  \label{C2}
\end{equation}

For the Minkowski bulk we have $c=0$ and, hence, $\nu =1/2$. In this case
the Macdonald function in Eq. (\ref{Amodes}) is reduced to the exponential
one and we obtain the mode functions in the Minkowski spacetime. The special
value $\nu =1/2$ is obtained for general $c$ in spatial dimensions $D=3$
and, again, the mode functions (\ref{Amodes}) reduce to the Minkowskian
modes. This property is a consequence of the conformal invariance of the
electromagnetic field in $D=3$.

\subsection{Two-point functions}

We consider a free field theory and all the properties of the vacuum state
are encoded in the two-point functions. Two-point functions for vector
fields in maximally symmetric spaces have been investigated in Ref. \cite%
{Alle86}. Though the background geometry under consideration is less
symmetric, as it will be seen below, closed analytical expressions are
obtained in this case as well. Expanding the field operator in terms of the
complete set of mode functions and using the commutation relations for the
annihilation and creation operators, we get the following mode sum formula
for the two-point function of the vector potential:%
\begin{equation}
\langle A_{l}(x)A_{m}(x^{\prime })\rangle _{0}=\sum_{\sigma =1}^{D-1}\int d%
\mathbf{k}\,A_{(\sigma \mathbf{k})l}(x)A_{(\sigma \mathbf{k})m}^{\ast
}(x^{\prime }),  \label{AlAmMode}
\end{equation}%
where and in what follows the Latin indices for tensors run over $1,2,\ldots
,D$, and $\langle \cdots \rangle _{0}$ corresponds to the expectation value
in the boundary-free FRW spacetime. Substituting the expression (\ref{Amodes}%
) for the mode functions and introducing the notation $\Delta \mathbf{z=z}-%
\mathbf{z}^{\prime }$, the following integral representation is obtained:
\begin{equation}
\langle A_{l}(x)A_{m}(x^{\prime })\rangle _{0}=\frac{4(\eta \eta ^{\prime
})^{\nu }}{(2\pi )^{D}a_{0}^{D-3}}\int d\mathbf{k}\,e^{i\mathbf{k}\cdot
\Delta \mathbf{z}}\left( \delta _{lm}-\frac{k_{l}k_{m}}{k^{2}}\right) K_{\nu
}(k\eta e^{s_{\tau }\pi i/2})K_{\nu }(k\eta ^{\prime }e^{-s_{\tau }\pi i/2}).
\label{AlAm}
\end{equation}

In the integral with the term $\delta _{lm}$, we first integrate over the
angular part of $\mathbf{k}$. The remaining integral over $k$ is convergent
at the lower limit, $k=0$, under the condition $|\nu |<D/2$. From this
condition, for the allowed values of the parameter $c$ one finds:%
\begin{equation}
\left\{ c<\frac{D+1}{2(D-1)}\right\} \cup \left\{ c>\frac{D-1}{2}\right\} .
\label{callowed}
\end{equation}%
In this range of powers, the integral over $k$ is evaluated by using the
formula from Ref. \cite{Prud86} and it is expressed in terms of the
associated Legendre function of the first kind. We write the final result in
terms of the hypergeometric function:%
\begin{eqnarray}
\int d\mathbf{k\,}e^{i\mathbf{k}\cdot \Delta \mathbf{z}}K_{\nu }(k\eta
e^{s_{\tau }\pi i/2})K_{\nu }(k\eta ^{\prime }e^{-s_{\tau }\pi i/2}) &=&%
\frac{\pi ^{(D+1)/2}}{2\left( \eta \eta ^{\prime }\right) ^{D/2}}\frac{%
\Gamma (D/2+\nu )\Gamma (D/2-\nu )}{\Gamma \left( (D+1)/2\right) }  \notag \\
&&\times F\left( \frac{D}{2}+\nu ,\frac{D}{2}-\nu ;\frac{D+1}{2};z\right) ,
\label{IntKK}
\end{eqnarray}%
where%
\begin{equation}
z=1+\frac{\left( \Delta \eta \right) ^{2}-|\Delta \mathbf{z}|^{2}}{4\eta
\eta ^{\prime }},  \label{z}
\end{equation}%
and $\Delta \eta =\eta -\eta ^{\prime }$. With this integral, the two-point
function is presented as%
\begin{eqnarray}
\langle A_{l}(x)A_{m}(x^{\prime })\rangle _{0} &=&\delta _{lm}\frac{(\eta
\eta ^{\prime })^{\nu -D/2}}{(4\pi )^{(D-1)/2}}\frac{\Gamma (D/2+\nu )\Gamma
(D/2-\nu )}{\Gamma \left( (D+1)/2\right) a_{0}^{D-3}}F\left( \frac{D}{2}+\nu
,\frac{D}{2}-\nu ;\frac{D+1}{2};z\right)   \notag \\
&&-\frac{4(\eta \eta ^{\prime })^{\nu }}{(2\pi )^{D}a_{0}^{D-3}}\int d%
\mathbf{k}\,e^{i\mathbf{k}\cdot \Delta \mathbf{z}}\frac{k_{l}k_{m}}{k^{2}}%
K_{\nu }(k\eta e^{s_{\tau }\pi i/2})K_{\nu }(k\eta ^{\prime }e^{-s_{\tau
}\pi i/2}).  \label{AlAm2}
\end{eqnarray}%
For the integral term in this formula one has no closed analytic expression.
However, this part of the two-point function will not contribute to the
two-point function for the field tensor and it will not be needed in the
evaluation of the VEVs for the field squared and the energy-momentum tensor.

Given the two-point function for the vector potential, the corresponding
function for the electromagnetic field tensor is evaluated as%
\begin{equation}
\langle F_{\mu \lambda }(x)F_{\gamma \delta }(x^{\prime })\rangle
_{0}=4\delta _{\lbrack \lambda }^{\rho }\partial _{\mu ]}\delta _{\lbrack
\delta }^{\sigma }\partial _{\gamma ]}^{\prime }\langle A_{\rho
}(x)A_{\sigma }(x^{\prime })\rangle _{0},  \label{FF0}
\end{equation}%
where the square brackets mean the antisymmetrization over the enclosed
indices: $a_{\cdots \lbrack i_{j}\cdots i_{k}]\cdots }=(a_{\cdots
i_{j}\cdots i_{k}\cdots }-a_{\cdots i_{k}\cdots i_{j}\cdots })/2$. The
second term in the right-hand side of Eq. (\ref{AlAm2}) does not contribute
and one gets the following expressions:%
\begin{eqnarray}
\langle F_{0l}(x)F_{0m}(x^{\prime })\rangle _{0} &=&\frac{(\eta \eta
^{\prime })^{\nu -D/2-1}}{2B_{D}a_{0}^{D-3}}\left[ \left( \delta _{lp}\delta
_{mq}-\delta _{lm}\delta _{pq}\right) \frac{\Delta z^{p}\Delta z^{q}}{2\eta
\eta ^{\prime }}\partial _{z}+\left( D-1\right) \delta _{lm}\right] F_{\nu
-1}(z),  \notag \\
\langle F_{0m}(x)F_{pl}(x^{\prime })\rangle _{0} &=&-s_{\tau }\frac{(\eta
\eta ^{\prime })^{\nu -D/2-1}}{B_{D}a_{0}^{D-3}}\delta _{\lbrack pm}\delta
_{l]q}\frac{\Delta z^{q}}{\eta }\left[ \nu -\frac{D}{2}-1+\left( \frac{\eta
+\eta ^{\prime }}{2\eta ^{\prime }}-z\right) \partial _{z}\right] F_{\nu
}(z),  \label{FF0m} \\
\langle F_{pl}(x)F_{qm}(x^{\prime })\rangle _{0} &=&\frac{(\eta \eta
^{\prime })^{\nu -D/2-1}}{B_{D}a_{0}^{D-3}}\left( \delta _{\lbrack pr}\delta
_{l][q}\delta _{m]s}\frac{\Delta z^{r}\Delta z^{s}}{\eta \eta ^{\prime }}%
\partial _{z}+2\delta _{\lbrack pq}\delta _{l]m}\right) F_{\nu }(z),  \notag
\end{eqnarray}%
with the notation%
\begin{equation}
B_{D}=(4\pi )^{(D-1)/2}\Gamma \left( (D+1)/2\right) .  \label{BD}
\end{equation}%
For the further convenience, in Eq. (\ref{FF0m}) we have introduced the
function%
\begin{equation}
F_{\nu }(z)=\frac{2}{D+1}\Gamma \left( \frac{D}{2}+1+\nu \right) \Gamma
\left( \frac{D}{2}+1-\nu \right) F\left( \frac{D}{2}+1+\nu ,\frac{D}{2}%
+1-\nu ;\frac{D+3}{2};z\right) .  \label{FG}
\end{equation}%
The expression for the component $\langle F_{pl}(x)F_{0m}(x^{\prime
})\rangle _{0}$ is obtained from that for $\langle F_{0m}(x)F_{pl}(x^{\prime
})\rangle _{0}$ by changing the sign and by the interchange $\eta
\rightleftarrows \eta ^{\prime }$. The two-point function for the field
tensor is obtained from the two-point function of the vector potential
taking derivatives with respect to the coordinates. This enlarges the region
for the values of the parameter $c$ for which the infrared divergences are
absent in the integral representation of the correlator $\langle F_{\mu
\lambda }(x)F_{\gamma \delta }(x^{\prime })\rangle _{0}$. Now, the infrared
divergences for the Bunch-Davies vacuum state are absent under the condition
$-D/2<\nu <D/2+1$ or, in terms of the parameter $c$,
\begin{equation}
\left\{ c<\frac{D+1}{2(D-1)}\right\} \cup \left\{ c>\frac{D+1}{4}\right\} .
\label{callowedF}
\end{equation}%
For the values of $c$ outside this range the two-point function for the
field tensor contains infrared divergences and the Bunch-Davies vacuum in a
boundary-free FRW background is not a physically realizable state. Note
that, in the limit $c\rightarrow \infty $ one has $\nu =D/2-1$ and from Eq. (%
\ref{FF0}) we obtain the corresponding two-point functions in de Sitter
spacetime with the scale factor $a(t)=e^{t/a_{0}}$ (see Ref. \cite{Saha13}).

Note that $F_{\nu }(z)$ is an even function of $\nu $: $F_{-\nu }(z)=F_{\nu
}(z)$. An equivalent form for this function is obtained by using the linear
transformation formula for the hypergeometric function \cite{Abra72}:
\begin{equation}
F_{\nu }(z)=\frac{2\Gamma \left( D/2+1-\nu \right) \Gamma \left( D/2+1+\nu
\right) }{\left( D+1\right) (1-z)^{(D+1)/2}}F\left( \frac{1}{2}-\nu ,\frac{1%
}{2}+\nu ;\frac{D+3}{2};z\right) .  \label{FG1}
\end{equation}%
Now, by taking into account that $F\left( 0,1;c;z\right) =1$, for $\nu =1/2$
we get%
\begin{equation}
F_{1/2}(z)=\frac{\Gamma ^{2}\left( (D+1)/2\right) }{(1-z)^{(D+1)/2}}.
\label{F12}
\end{equation}%
As we have already mentioned before, the value $\nu =1/2$ is obtained in two
special cases: (i) for $D=3$ with general $c$ and (ii) for $c=0$. The latter
corresponds to the case of the Minkowski bulk. Hence, taking in Eq. (\ref%
{FF0m}) $\nu =1/2$, $a_{0}=1$, with the function $F_{1/2}(z)$ from Eq. (\ref%
{F12}), we obtain the corresponding two-point functions in $(D+1)$%
-dimensional Minkowski spacetime.

In what follows we need the asymptotic expressions for the function $F_{\nu
}(z)$ when the argument is close to 1 and when it is large. These limits
correspond to small and large separations of the points in the arguments of
the two-point functions. For $|1-z|\ll 1$, by using the linear
transformation formula for the hypergeometric function (see, for example,
Ref. \cite{Abra72}), it can be seen that one has the asymptotic expression%
\begin{equation}
F_{\nu }(z)\approx \frac{\Gamma ^{2}((D+1)/2)}{(1-z)^{(D+1)/2}}\left[ 1-%
\frac{1-4\nu ^{2}}{2(D-1)}\left( 1-z\right) \right] .  \label{FnuAs1}
\end{equation}%
Note that the leading term in this formula coincides with Eq. (\ref{F12}).
By taking into account that the latter corresponds to the Minkowski
spacetime, we conclude that the leading divergence of the two-point function
in the coincidence limit of the arguments coincides with that for the
Minkowski spacetime. This is a consequence of our choice of the vacuum
state. For $-z\gg 1$, again, with the help of the linear transformation
formula for the hypergeometric function, for $\nu \neq 0$ the following
asymptotic expression can be obtained for the function $F_{\nu }(z)$:%
\begin{equation}
F_{\nu }(z)\approx \frac{2^{2|\nu |-1}\Gamma ((D+1)/2)}{\sqrt{\pi }%
(-z)^{D/2+1-|\nu |}}\Gamma (|\nu |)\Gamma \left( D/2-|\nu |+1\right) .
\label{Fas2}
\end{equation}%
In the case $\nu =0$, the asymptotic expression is obtained by using the
formula 15.3.14 from Ref. \cite{Abra72}:%
\begin{equation}
F_{0}(z)\approx \frac{\Gamma (D+1)}{2^{D}(-z)^{D/2+1}}\ln (-z),\;-z\gg 1.
\label{F0as2}
\end{equation}%
Note that $\nu =0$ corresponds to $c=1/(D-2)$. For $D>3$ this falls into the
range (\ref{callowed}) for the allowed values of the parameter $c$. For $D>3$
and $0<c<2/(D-1)$ the decay of the two-point function at large separations
of the points is faster than in the Minkowskian case. For other values of $c$
allowed by Eq. (\ref{callowed}), the decay is slower and the correlation of
the vacuum fluctuations is stronger. In particular, the latter is the case
in power law inflationary models.

\section{Geometry of two conducting plates}

\label{sec:TPpl}

In this section we consider two perfectly conducting plates in background of
the FRW spacetime described by the line element (\ref{ds2}), assuming that
they are placed at $z^{D}=0$ and $z^{D}=L$. On the plates the field obeys
the boundary condition \cite{Ambj83}
\begin{equation}
n^{\nu _{1}}\,^{\ast }F_{\nu _{1}\cdots \nu _{D-1}}=0,  \label{BC}
\end{equation}%
with the tensor $^{\ast }F_{\nu _{1}\cdots \nu _{D-1}}$ dual to $F_{\mu \nu
} $, and $n^{\mu }$ is the normal to the plates. For the coordinates and the
momentum components parallel to the plates we shall use the notations $%
\mathbf{z}_{\parallel }=(z^{1},\ldots ,z^{D-1})$ and $\mathbf{k}_{\parallel
}=(k_{1},\ldots ,k_{D-1})$. In the region between the plates, for the mode
functions, obeying the gauge conditions and the boundary condition on the
plate $z^{D}=0$, one has%
\begin{eqnarray}
A_{(\sigma \mathbf{k})l}(x) &=&iC\epsilon _{(\sigma )l}\eta ^{\nu }K_{\nu
}(k\eta e^{s_{\tau }\pi i/2})\sin (k_{D}x^{D})e^{i\mathbf{k}_{\parallel
}\cdot \mathbf{z}_{\parallel }},  \notag \\
A_{(\sigma \mathbf{k})D}(x) &=&C\epsilon _{(\sigma )D}\eta ^{\nu }K_{\nu
}(k\eta e^{s_{\tau }\pi i/2})\cos (k_{D}x^{D})e^{i\mathbf{k}_{\parallel
}\cdot \mathbf{z}_{\parallel }},  \label{Amode}
\end{eqnarray}%
where $l=1,\ldots ,D-1$, and $k=\sqrt{k_{D}^{2}+\mathbf{k}_{\parallel }^{2}}$%
. For the polarization vector $\epsilon _{(\sigma )l}$ we have the same
relations as before (see Eqs. (\ref{Eps1}) and (\ref{Eps2})). From the
boundary condition at $z^{D}=L$ for the eigenvalues of $k_{D}$ we get$\ $%
\begin{equation}
k_{D}=\pi n/L,\;n=0,1,\ldots  \label{kDval}
\end{equation}%
Now the normalization condition is given by%
\begin{equation}
\int d\mathbf{z}_{\Vert }\int_{0}^{L}dz^{D}\sum_{l=1}^{D}[A_{(\sigma \mathbf{%
k})l}^{\ast }(x)\partial _{\tau }\ A_{(\sigma ^{\prime }\mathbf{k}^{\prime
})l}(x)-A_{(\sigma ^{\prime }\mathbf{k}^{\prime })_{l}}(x)\partial _{\tau }\
A_{(\sigma \mathbf{k})l}^{\ast }(x)]=\frac{4\pi \delta _{\sigma \sigma
^{\prime }}\delta _{nn^{\prime }}}{ia^{D-3}(\tau )}\delta (\mathbf{k}%
_{\parallel }-\mathbf{k}_{\parallel }^{\prime }).  \label{NormCond}
\end{equation}%
From this condition for the coefficient $C$ in Eq. (\ref{Amode}) one finds%
\begin{equation*}
\ \left\vert C\right\vert ^{2}=\frac{a_{0}^{3-D}\delta _{n}}{2(2\pi )^{D-3}L}%
,
\end{equation*}%
with $\delta _{n}=1$ for $n\neq 0$ and $\delta _{0}=1/2$.

Substituting the mode functions into the mode sum formula for the two-point
function of the vector potential, similar to Eq. (\ref{AlAmMode}), one finds
the following integral representation%
\begin{eqnarray}
\langle A_{l}(x)A_{m}(x^{\prime })\rangle &=&\frac{2a_{0}^{3-D}\left( \eta
\eta ^{\prime }\right) ^{\nu }}{(2\pi )^{D-1}L}\int d\mathbf{k}_{\parallel
}e^{i\mathbf{k}_{\parallel }\cdot \Delta \mathbf{z}_{\parallel
}}\sum_{n=-\infty }^{\infty }\left( \delta _{lm}-\frac{k_{l}k_{m}}{%
k_{(n)}^{2}}\right)  \notag \\
&&\times K_{\nu }(k_{(n)}\eta e^{s_{\tau }\pi i/2})K_{\nu }(k_{(n)}\eta
^{\prime }e^{-s_{\tau }\pi i/2})\sum_{j=\pm 1}j^{1-\delta
_{mD}}e^{ik_{D}\left( z^{D}-jz^{D\prime }\right) },  \label{AlAm2pl}
\end{eqnarray}%
where $k_{(n)}=\sqrt{(\pi n/L)^{2}+\mathbf{k}_{\parallel }^{2}}$. Note that
for the terms $n\neq 0$ one has $k_{(n)}\neq 0$ and in the integrals with
these terms there are no infrared divergences for all values of $\nu $. The
only restriction on the values of this parameter comes from the requirement
of the convergence of the integral with the term $n=0$ at $k_{(0)}=|\mathbf{k%
}_{\parallel }|$: $|\nu |<(D-1)/2$ or, in terms of $c$, $\left\{
c<D/(2D-3)\right\} \cup \left\{ c>D-2\right\} $.

By applying the Poisson resummation formula in the form%
\begin{equation}
\sum_{n=-\infty }^{+\infty }g(\pi n/L)=\frac{L}{\pi }\sum_{n=-\infty
}^{+\infty }\int_{-\infty }^{+\infty }dy\,e^{-2inLy}g(y),  \label{Poiss}
\end{equation}%
the two-point function (\ref{AlAm2pl}) is expressed in terms of the
corresponding function in the boundary-free geometry:%
\begin{equation}
\langle A_{l}(x)A_{m}(x^{\prime })\rangle =\sum_{n=-\infty }^{\infty
}\sum_{j=\pm }j^{1-\delta _{mD}}\langle A_{l}(x)A_{m}(x_{j,n}^{\prime
})\rangle _{0},  \label{AlAm2pl1}
\end{equation}%
where%
\begin{equation}
x_{\pm ,n}^{\prime }=\left( \tau ^{\prime },z^{\prime 1},\ldots ,\pm
z^{\prime D}+2nL\right) .  \label{xplmi}
\end{equation}%
The two-point function for the electromagnetic field tensor is directly
obtained from Eq. (\ref{AlAm2pl1}):%
\begin{equation}
\langle F_{pl}(x)F_{qm}(x^{\prime })\rangle =\sum_{n=-\infty }^{+\infty
}\sum_{j=\pm }j^{1-\delta _{qD}}\langle F_{pl}(x)F_{qm}(x_{j,n}^{\prime
})\rangle _{0}.  \label{FF2pl}
\end{equation}%
Note that in Eqs. (\ref{AlAm2pl1}) and (\ref{FF2pl}), the term $n=0$, $j=+1$
presents the corresponding two-point function in the boundary-free geometry
and the term $n=0$, $j=-1$ is the part in the two-point function induced by
a plate at $z^{D}=0$ when the right plate is absent. The integrals in the
integral representation of the correlator $\langle F_{pl}(x)F_{qm}(x^{\prime
})\rangle $ are infrared convergent under the condition $-(D-1)/2<\nu
<(D+1)/2$. In terms of the parameter $c$ this constraint is written as
\begin{equation}
\left\{ c<D/(2D-3)\right\} \cup \left\{ c>D/3\right\} .  \label{callowed2}
\end{equation}%
This condition will be assumed below in the evaluation of the VEVs for the
field squared and the energy-momentum tensor.

We refer to the boundaries with the condition (\ref{BC}) as conducting
plates. In $D=3$ this type of boundary condition is realized by perfect
conductors. In a background with $D>3$, the plates, in general, are just
hypersurfaces (for example, branes in Randall-Sundrum-type models)
reflecting the modes of the field. Instead we could consider the boundary
condition (infinitely permeable boundary condition) $n^{\mu }F_{\mu \beta
}=0 $, which is obtained by requiring the action for the field to vanish
outside a bounded region (see Ref. \cite{Ambj83}). This condition for gluon
fields is used in bag models for hadrons. The two-point functions and the
VEVs for the case of infinitely permeable boundary condition are considered
in a way similar to that we describe here for conducting plates.

\section{Casimir densities in the geometry of a single plate}

\label{sec:Cas1pl}

The VEVs of the field squared and the energy-momentum tensor are among the
most important characteristics of the vacuum state. In this section we
investigate these VEVs in the geometry of a single conducting plate placed
at $z^{D}=0$. The corresponding two-point function for the field tensor
(denoted below by the index 1) is given by the $n=0$ term in Eq. (\ref{FF2pl}%
). This function is presented in the decomposed form
\begin{equation}
\langle F_{pl}(x)F_{qm}(x^{\prime })\rangle _{1}=\langle
F_{pl}(x)F_{qm}(x^{\prime })\rangle _{0}+\langle F_{pl}(x)F_{qm}(x^{\prime
})\rangle _{1b},  \label{FF1pl}
\end{equation}%
where the second term in the right-hand is induced by the presence of the
plate and is given by the expression
\begin{equation}
\langle F_{pl}(x)F_{qm}(x^{\prime })\rangle _{1b}=(-1)^{1-\delta
_{qD}}\langle F_{pl}(x)F_{qm}(x_{-,0}^{\prime })\rangle _{0}.
\label{FF1pl1b}
\end{equation}%
Here, $x_{-,0}^{\prime }=\left( \tau ^{\prime },z^{\prime 1},\ldots
,-z^{\prime D}\right) $ is the image of the spacetime point $x^{\prime
}=\left( \tau ^{\prime },z^{\prime 1},\ldots ,z^{\prime D}\right) $ with
respect to the plate. Given the two-point function, we can evaluate the VEVs
of the field squared and the energy-momentum tensor. We start with the
electric field squared.

\subsection{Field squared}

For the VEV of the electric field squared one has%
\begin{equation}
\langle E^{2}\rangle _{1}=-g^{00}g^{lm}\lim_{x^{\prime }\rightarrow
x}\langle F_{0l}(x)F_{0m}(x^{\prime })\rangle _{1}.  \label{E2}
\end{equation}%
The expression in the right-hand side is divergent and in order to extract a
finite physical result some renormalization procedure is necessary. An
important point here is that for $z^{D}\neq 0$ the conducting plate does not
change the local geometry of the background spacetime and, hence, the
divergences in the coincidence limit are the same as in the boundary-free
geometry. We have already decomposed the two-point functions into the
boundary-free and plate-induced parts. For points away from the plate the
divergences are contained in the boundary-free part only and the
renormalized VEV is decomposed as
\begin{equation}
\langle E^{2}\rangle _{1}=\langle E^{2}\rangle _{0}+\langle E^{2}\rangle
_{1b},  \label{E21}
\end{equation}%
where $\langle E^{2}\rangle _{0}$ is the renormalized VEV of the field
squared in the absence of the plate. The background geometry is homogeneous
and this VEV does not depend on the spatial point. In the present paper we
are interested in the effects induced by the plate.

The boundary-induced part in Eq. (\ref{E21}) is finite for points away from
the plate and it is directly evaluated by using Eqs. (\ref{FF0m}) and (\ref%
{FF1pl1b}). This gives%
\begin{equation}
\langle E^{2}\rangle _{1b}=\frac{D-1}{2B_{D}(\eta a)^{D+1}}\left[
2(1-y)\partial _{y}-D+2\right] F_{\nu -1}(y),  \label{E1pl1b}
\end{equation}%
where and in what follows we use the notation%
\begin{equation}
y=1-\left( z^{D}/\eta \right) ^{2}.  \label{yv}
\end{equation}%
The plate-induced part depends on the coordinate $z^{D}$ in the combination $%
z^{D}/\eta $. We consider the region $z^{D}>0$ and the proper distance from
the plate is given by $az^{D}$. Note that $z^{D}/\eta \propto az^{D}/t$. By
taking into account that the curvature radius of the background spacetime is
proportional to $t$, we conclude that, up to a constant factor, $z^{D}/\eta $
is the ratio of the proper distance from the plate to the curvature radius
of the background spacetime. For a fixed $z^{D}/\eta $, the dependence on
time appears in Eq. (\ref{E1pl1b}) in the form of the product $\eta a$. The
latter is expressed in terms of the comoving time coordinate and the Hubble
function, $H=a^{-1}\partial _{t}a$, as%
\begin{equation}
\eta a=\frac{|t|}{|1-c|}=\frac{|b|}{H}.  \label{etaH}
\end{equation}%
Now, we see that the VEV of the field squared, multiplied by the Hubble
volume $H^{-D}$, has the functional form $H^{-D}\langle E^{2}\rangle
_{1b}=Hf(z^{D}/\eta )$.

In the special case $\nu =1/2$, by using Eq. (\ref{F12}), from Eq. (\ref%
{E1pl1b}) we find
\begin{equation}
\langle E^{2}\rangle _{1b}=\frac{3(D-1)\Gamma \left( (D+1)/2\right) }{2(4\pi
)^{(D-1)/2}(az^{D})^{D+1}}.  \label{E21M}
\end{equation}%
For $a=1$, this gives the VEV of the field squared for a conducting plate in
$(D+1)$-dimensional Minkowski spacetime. For $D=3$, from Eq. (\ref{E21M})
one gets a simple result $\langle E^{2}\rangle _{1b}=3\left( az^{D}\right)
^{-4}/(4\pi )$. The electromagnetic field is conformally invariant in $D=3$
and this result is directly obtained from the corresponding expression in
the Minkowski spacetime by a conformal transformation.

Simple expressions for the plate-induced part in the VEV of the electric
field squared are obtained near the plate and at large distances. For points
near the plate one has $z^{D}/\eta \ll 1$ and the argument of the function $%
F_{\nu -1}(y)$ in Eq. (\ref{E1pl1b}) is close to 1. By using the asymptotic
formula (\ref{FnuAs1}), to the leading order, from Eq. (\ref{E1pl1b}) we get%
\begin{equation}
\langle E^{2}\rangle _{1b}\approx \frac{3(D-1)\Gamma ((D+1)/2)}{2(4\pi
)^{(D-1)/2}(az^{D})^{D+1}}.  \label{E1plAs1}
\end{equation}%
This leading term coincides with the corresponding VEV for a plate in the
Minkowski spacetime with the distance from the plate replaced by $az^{D}$.
As it is seen, near the plate the boundary-induced part in the VEV of the
electric field squared is positive and diverges on the plate. By taking into
account that the boundary-free part does not depend on the spatial point, we
conclude that near the plate $\langle E^{2}\rangle _{1b}$ dominates in Eq. (%
\ref{E21}) and the total VEV is positive as well. The surface divergences in
the VEVs of local physical observables are well known in quantum field
theory with boundaries (see, for instance, Ref. \cite{Eliz94}). These
divergences are related to the hard boundary conditions imposed on modes of
all wavelengths.

At distances from the plate much larger than the curvature radius of the
background spacetime one has $z^{D}/\eta \gg 1$ and, hence, $-y\gg 1$. By
using the asymptotic expressions (\ref{Fas2}) and (\ref{F0as2}), for the
plate-induced part in the VEV of the electric field squared, in the leading
order, we get%
\begin{equation}
\langle E^{2}\rangle _{1b}\approx \frac{2^{2|\nu -1|}\left( D-1\right)
\left( 2-|\nu -1|\right) \Gamma (|\nu -1|)}{(4\pi )^{D/2}(\eta
a)^{D+1}(z^{D}/\eta )^{D+2-2|\nu -1|}}\Gamma \left( D/2-|\nu -1|+1\right) ,
\label{E1plAs2}
\end{equation}%
for $\nu \neq 1$ and%
\begin{equation}
\langle E^{2}\rangle _{1b}\approx 8\left( D-1\right) \frac{\Gamma \left(
D/2+1\right) }{(4\pi )^{D/2}}\frac{\eta a\ln (z^{D}/\eta )}{(az^{D})^{D+2}},
\label{E1plAs2b}
\end{equation}%
for $\nu =1$. The latter corresponds to $c=1/(4-D)$. For $\nu =-1$ and $\nu
=3$ the leading term (\ref{E1plAs2}) vanishes and we need to keep the next
to the leading order contribution. At large distances from the plate the
boundary-induced part tends to zero and the total VEV of the electric field
squared is dominated by the boundary-free part. From Eq. (\ref{E1plAs2}) it
follows that, at large distances from the plate one has $\langle
E^{2}\rangle _{1b}>0$ for $-1<\nu <3$. In terms of the parameter $c$, we can
see that in the case $4\leqslant D<8$, at large distances the plate-induced
part is positive for $\{c<3/D\}\cup \{c>5/(8-D)\}$. For $D>8$ one has $%
\langle E^{2}\rangle _{1b}>0$ for $5/(8-D)<c<3/D$. In the case $D=8$, the
plate-induced part is positive for $c<3/8$. As we see, there is a region for
the values of the parameter $c$ in which the plate-induced part in the VEV
of the electric field squared is positive near the plate and negative at
large distances.

In the geometry of a single conducting plate, we have a similar
decomposition for the Lagrangian density. The part in the corresponding VEV
induced by the plate is given as
\begin{equation}
\langle F_{\beta \sigma }F^{\beta \sigma }\rangle _{1b}=-2\langle
E^{2}\rangle _{1b}-\frac{D-1}{B_{D}(\eta a)^{D+1}}\left[ 2(1-y)\partial
_{y}+D-4\right] F_{\nu }(y).  \label{Lag1b}
\end{equation}%
The second term in the right-hand side presents the magnetic contribution.
In $D=3$ it coincides with the first term. The asymptotics of the magnetic
part are investigated in a way similar to that we have described for the
case of the electric field squared.

\subsection{Energy-momentum tensor}

Having the two-point function for the field tensor we can evaluate the VEV\
of the energy-momentum tensor by using the formula
\begin{equation}
\langle T_{\mu }^{\nu }\rangle =-\frac{1}{4\pi }\lim_{x^{\prime }\rightarrow
x}\langle F_{\mu }^{\cdot \beta }(x)F_{\cdot \beta }^{\nu }(x^{\prime
})\rangle +\frac{\delta _{\mu }^{\nu }}{16\pi }\langle F_{\beta \sigma
}F^{\beta \sigma }\rangle .  \label{Tb}
\end{equation}%
By making use of Eq. (\ref{FF1pl}), in the geometry of a single conducting
plate this VEV is decomposed as%
\begin{equation}
\langle T_{\mu }^{\nu }\rangle _{1}=\langle T_{\mu }^{\nu }\rangle
_{0}+\langle T_{\mu }^{\nu }\rangle _{1b},  \label{T1dec}
\end{equation}%
where, for points outside the plate, the renormalization is required for the
boundary-free part $\langle T_{\mu }^{\nu }\rangle _{0}$ only. Because of
the spatial isotropy of the background spacetime the corresponding stresses
are isotropic, $\langle T_{1}^{1}\rangle _{0}=\cdots =\langle
T_{D}^{D}\rangle _{0}$, and one has only two algebraically independent
components, $\langle T_{0}^{0}\rangle _{0}$ and $\langle T_{D}^{D}\rangle
_{0}$. These components must be functions of the time only due to the
spatial homogeneity. A differential relation between two components is
obtained from the covariant conservation equation $\nabla _{\nu }\langle
T_{\mu }^{\nu }\rangle _{0}=0$. In $D=3$ the electromagnetic field is
conformally invariant and the VEV $\langle T_{\mu }^{\nu }\rangle _{0}$ is
completely determined by the trace anomaly (see, for instance, Ref. \cite%
{Birr82}).

For the plate-induced parts in the VEVs for the diagonal components of the
energy-momentum tensor we get the following expressions (there is no sum on $%
\mu $)%
\begin{equation}
\langle T_{\mu }^{\mu }\rangle _{1b}=\frac{C_{D}f_{\nu }^{(\mu )}(y)}{2(\eta
a)^{D+1}},  \label{Tmu1b}
\end{equation}%
with the notations ($l=1,\ldots ,D-1$)%
\begin{eqnarray}
f_{\nu }^{(0)}(y) &=&\left[ 2\left( 1-y\right) \partial _{y}-D+2\right] %
\left[ F_{\nu -1}(y)-F_{\nu }(y)\right] -2\left( D-3\right) F_{\nu }(y),
\notag \\
f_{\nu }^{(l)}(y) &=&2\frac{3-D}{D-1}\left\{ \left( 1-y\right) \partial _{y}%
\left[ F_{\nu }(y)+F_{\nu -1}(y)\right] -2F_{\nu }(y)\right\} -\left(
D-4\right) \left[ F_{\nu }(y)-F_{\nu -1}(y)\right] ,  \label{fly} \\
f_{\nu }^{(D)}(y) &=&\left[ 2\left( 1-y\right) \partial _{y}-D\right] \left[
F_{\nu }(y)-F_{\nu -1}(y)\right] ,  \notag
\end{eqnarray}%
and%
\begin{equation}
C_{D}=\frac{(4\pi )^{-(D+1)/2}}{\Gamma \left( (D-1)/2\right) }.  \label{CD}
\end{equation}%
In addition to the diagonal components, the vacuum average of the
energy-momentum tensor has also a nonzero off-diagonal component
\begin{equation}
\langle T_{0}^{D}\rangle _{1b}=-\frac{2s_{\tau }C_{D}}{(\eta a)^{D+1}}\frac{%
z^{D}}{\eta }f^{(0D)}(y),  \label{T0D1b}
\end{equation}%
with%
\begin{equation}
f_{\nu }^{(0D)}(y)=\left[ (1-y)\partial _{y}+\nu -D/2-1\right] F_{\nu }(y).
\label{f0D}
\end{equation}%
This component describes the energy flux along the direction normal to the
plate. Recall that the boundary-free part in the VEV of the energy-momentum
tensor is diagonal and the energy flux is purely boundary-induced effect.
Similar to the case of the field squared, the combination $H^{-(D+1)}\langle
T_{\mu }^{\nu }\rangle $, with $H$ been the Hubble parameter, is a function
of the ratio $z^{D}/\eta $ only.

In the special case $\nu =1/2$ we have the following simple expression (no
summation over $\mu $)%
\begin{equation}
\langle T_{\mu }^{\mu }\rangle _{1b}=\frac{(D-1)\left( 3-D\right) \Gamma
\left( (D+1)/2\right) }{2(4\pi )^{(D+1)/2}\left( az^{D}\right) ^{D+1}},
\label{TmuM}
\end{equation}%
for $\mu =0,1,\ldots ,D-1$, and $\langle T_{D}^{D}\rangle _{1b}=\langle
T_{0}^{D}\rangle _{1b}=0$. In particular, for $a=1$ the formula (\ref{TmuM})
gives the VEV in the Minkowski spacetime. Note that in this case the
parallel stresses are equal to the energy density. Of course, this property
is a direct consequence of the invariance of the problem with respect to the
Lorentz boosts along the directions parallel to the plate. For the plate in
FRW spacetime the parallel stresses, in general, differ from the energy
density. For $D=3$ the plate-induced contribution in the VEV of the
energy-momentum tensor vanishes. The electromagnetic field is conformally
invariant in $D=3$ and this result could also be directly obtained from the
corresponding result in the Minkowski spacetime by conformal transformation.

The plate-induced part separately obeys the covariant conservation equation $%
\nabla _{\nu }\langle T_{\mu }^{\nu }\rangle _{1b}=0$. For the problem under
consideration this reduces to the equations%
\begin{eqnarray}
\left( \partial _{\eta }+\frac{Db}{\eta }\right) \langle T_{0}^{0}\rangle
_{1b}+s_{\tau }\partial _{D}\langle T_{0}^{D}\rangle _{1b}-\frac{b}{\eta }%
\sum_{l=1}^{D}\langle T_{l}^{l}\rangle _{1b} &=&0,  \notag \\
\left( \partial _{\eta }+\frac{D+1}{\eta }b\right) \langle T_{0}^{D}\rangle
_{1b}-s_{\tau }\partial _{D}\langle T_{D}^{D}\rangle _{1b} &=&0,
\label{ContEq}
\end{eqnarray}%
with the parameter $b$ defined in Eq. (\ref{b}). The second equation in Eq. (%
\ref{ContEq}) relates the off-diagonal component in the VEV with the $z^{D}$%
-dependence of the normal stress. The expressions (\ref{Tmu1b}) and (\ref%
{T0D1b}) provide the components of the vacuum energy-momentum tensor in the
coordinate system $(\tau ,z^{1},\ldots ,z^{D})$. Let $\langle T_{(c)\mu
}^{\nu }\rangle _{1b}$ be the corresponding components in the coordinate
system with the comoving time, $(t,z^{1},\ldots ,z^{D})$. We have simple
relations: $\langle T_{(c)\mu }^{\mu }\rangle _{1b}=\langle T_{\mu }^{\mu
}\rangle _{1b}$ (no sum over $\mu =0,\ldots ,D$) and $\langle
T_{(c)0}^{D}\rangle _{1b}=a^{-1}(t)\langle T_{0}^{D}\rangle _{1b}$. For the
plate-induced part of the vacuum energy in the volume $V$ with the boundary $%
\partial V$, measured by a comoving observer, one has $E_{V}^{(1b)}=a^{D}%
\int_{V}dz^{D}\,\langle T_{(c)0}^{0}\rangle _{1b}$. From the covariant
continuity equation for $\langle T_{(c)\mu }^{\nu }\rangle _{1b}$ the
following relation is obtained%
\begin{equation}
\partial _{t}E_{V}^{(1b)}=-\int_{\partial V}d^{D-1}z\,\sqrt{h}n_{l}\langle
T_{(c)0}^{l}\rangle _{1b}+\frac{c}{t}\int_{V}d^{D}z\sqrt{\gamma }\langle
T_{(c)l}^{l}\rangle _{1b},  \label{En}
\end{equation}%
where $n_{l}$, $\gamma ^{il}n_{i}n_{l}=1$, is the external normal to the
boundary $\partial V$, $\gamma _{il}=-g_{il}$ is the spatial metric tensor, $%
\gamma =\mathrm{det}(\gamma _{il})$, and $h$ is the determinant of the
induced metric $h_{il}=\gamma _{il}-n_{i}n_{l}$. The first term in the
right-hand side of Eq. (\ref{En}) corresponds to the energy flux through the
boundary. In particular, the energy flux through the surface at $z^{D}=%
\mathrm{const}$ is given by $a^{D}S\langle T_{(c)0}^{D}\rangle _{1b}$, where
$S=\int dz^{1}\cdots dz^{D-1}$ is the coordinate surface area. Now, by
taking into account that the proper area is given by $a^{D-1}S$, we conclude
that $a\langle T_{(c)0}^{D}\rangle _{1b}=\langle T_{0}^{D}\rangle _{1b}$ is
the energy flux per unit proper surface area.

Now we turn to the investigation of the asymptotic behavior of the
plate-induced parts (\ref{Tmu1b}) and (\ref{T0D1b}) near the plate and at
large proper distances from it compared to the curvature radius of the
background spacetime. At small distances, which correspond to $z^{D}/\eta
\ll 1$, we use the asymptotic expression (\ref{FnuAs1}) with the results (no
sum over $\mu =0,\ldots ,D-1$)%
\begin{eqnarray}
\langle T_{\mu }^{\mu }\rangle _{1b} &\approx &-\frac{\left( D-3\right)
(D-1)\Gamma ((D+1)/2)}{2(4\pi )^{(D+1)/2}(az^{D})^{D+1}},  \notag \\
\langle T_{D}^{D}\rangle _{1b} &\approx &\frac{(1-2\nu )\Gamma ((D+1)/2)}{%
2(4\pi )^{(D+1)/2}(\eta a)^{2}(az^{D})^{D-1}}.  \label{TDDas1}
\end{eqnarray}%
For the off-diagonal component we get%
\begin{equation}
\langle T_{0}^{D}\rangle _{1b}\approx \frac{(D-1)|1-2\nu |\Gamma ((D+1)/2)}{%
2(4\pi )^{(D+1)/2}\eta a(az^{D})^{D}}.  \label{T0Das1}
\end{equation}%
Note that the leading terms in the energy density and in the parallel
stresses coincide with the corresponding expressions for the VEVs in the
Minkowski spacetime, where the distance from the plate is replaced by the
proper distance $az^{D}$ (see Eq. (\ref{TmuM})). Near the plate the
boundary-induced part is dominant and the expressions (\ref{TDDas1}) and (%
\ref{T0Das1}) provide the leading terms in the asymptotic expansion of the
total VEV. As it is seen from Eq. (\ref{TDDas1}), the energy density and the
parallel stresses are negative near the plate for $D>3$. The normal stress
is positive for $0<c<1$ (decelerated expansion) and negative for $c<0$ or $%
c>1$ (accelerated expansion). For $D>3$, the off-diagonal component is
positive near the plate, $\langle T_{0}^{D}\rangle _{1b}>0$, and the energy
flux is directed from the plate for both decelerated and accelerated
expansions.

At large distances from the plate, $z^{D}/\eta \gg 1$, and for $\nu <1/2$,
in the expressions (\ref{fly}) the terms with the function $F_{\nu -1}(y)$
dominate. By using the asymptotic formula (\ref{Fas2}) for this function, to
the leading order we get (no summation over $\mu $)
\begin{equation}
\langle T_{\mu }^{\mu }\rangle _{1b}\approx q_{\nu }^{(\mu )}\frac{2^{1-2\nu
}(D-1)\Gamma (1-\nu )\Gamma \left( D/2+\nu \right) }{(4\pi )^{D/2+1}(\eta
a)^{D+1}(z^{D}/\eta )^{D+2\nu }},  \label{Tlarge1}
\end{equation}%
where%
\begin{equation}
q_{\nu }^{(0)}=\nu +1,\;q_{\nu }^{(l)}=-\frac{\left( D-3\right) \nu +D-2}{D-1%
},\;q_{\nu }^{(D)}=-\nu ,  \label{qunu}
\end{equation}%
with $l=1,\ldots ,D-1$. This case corresponds to $0<c<1$ with decelerating
expansion. For $D>3$, at large distances the plate-induced part in the VEV
of the energy density is positive for $c<3/D$ and negative for $c>3/D$. For $%
\nu >1/2$ and at large distances, the contribution of the terms with the
function $F_{\nu }(y)$ dominates and one has the following asymptotic
behavior (no summation over $\mu $):%
\begin{equation}
\langle T_{\mu }^{\mu }\rangle _{1b}\approx p_{\nu }^{(\mu )}\frac{2^{2\nu
-1}(D-1)\Gamma (\nu )\Gamma \left( D/2-\nu +1\right) }{(4\pi )^{D/2+1}(\eta
a)^{D+1}(z^{D}/\eta )^{D+2-2\nu }},  \label{Tlarge2}
\end{equation}%
where%
\begin{equation}
p_{\nu }^{(0)}=\nu +1-D,\;p_{\nu }^{(l)}=-\frac{\left( \nu +1\right) \left(
3-D\right) +D\left( D-4\right) +2}{D-1},\;\;p_{\nu }^{(D)}=1-\nu ,
\label{pnu}
\end{equation}%
and $l=1,\ldots ,D-1$. For this case one has $c<0$ or $c>1$ and the
expansion is accelerated. At large distances the plate-induced part in the
VEV of the energy density is positive for $1<c<2-3/D$. In both cases of the
decelerated and accelerated expansions the decay of the plate-induced
contributions in the energy density and in parallel stresses is weaker than
in the corresponding problem on Minkowski background. Note that the
asymptotic expressions are derived under the condition $(z^{D}/\eta )^{|2\nu
-1|}\gg 1$ they are not valid for $\nu $ too close to $1/2$.

In the consideration of the asymptotic behavior of the energy flux at large
distances from the plate the leading term is cancelled and we need to keep
in the asymptotic formula (\ref{Fas2}) the next to the leading term. Doing
that, for the function $f_{\nu }^{(0D)}(y)$ in the expression of the energy
flux to the leading order we get%
\begin{eqnarray}
f_{\nu }^{(0D)}(y) &\approx &-\frac{2^{-2\nu }\Gamma ((D+1)/2)}{\sqrt{\pi }%
(-y)^{D/2+1+\nu }}\Gamma (1-\nu )\Gamma \left( D/2+1+\nu \right) ,\;\nu <1/2,
\notag \\
f_{\nu }^{(0D)}(y) &\approx &\frac{2^{2\nu -2}\Gamma ((D+1)/2)}{\sqrt{\pi }%
(-y)^{D/2+2-\nu }}\Gamma (\nu )\Gamma \left( D/2+2-\nu \right) ,\;\nu >1/2,
\label{fnuas}
\end{eqnarray}%
where $-y\gg 1$. For the energy flux this gives
\begin{eqnarray}
\langle T_{0}^{D}\rangle _{1b} &\approx &\frac{2^{1-2\nu }(D-1)\Gamma (1-\nu
)\Gamma \left( D/2+1+\nu \right) }{(4\pi )^{D/2+1}(\eta a)^{D+1}(z^{D}/\eta
)^{D+1+2\nu }},\;\nu <1/2,  \label{T0Dlarge1} \\
\langle T_{0}^{D}\rangle _{1b} &\approx &\frac{2^{2\nu -1}(D-1)\Gamma (\nu
)\Gamma \left( D/2+2-\nu \right) }{(4\pi )^{D/2+1}(\eta a)^{D+1}(z^{D}/\eta
)^{D+3-2\nu }},\;\nu >1/2.  \label{T0Dlarge}
\end{eqnarray}%
In both cases, $\langle T_{0}^{D}\rangle _{1b}>0$ and at large distances the
energy flux is directed from the plate.

\begin{figure}[tbph]
\begin{center}
\begin{tabular}{cc}
\epsfig{figure=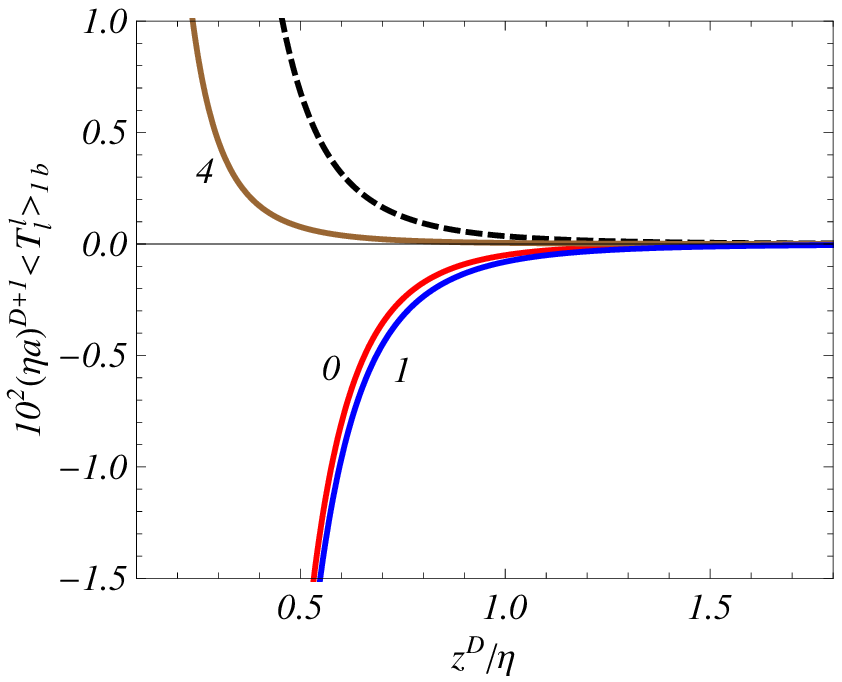,width=7.cm,height=6.cm} & \quad %
\epsfig{figure=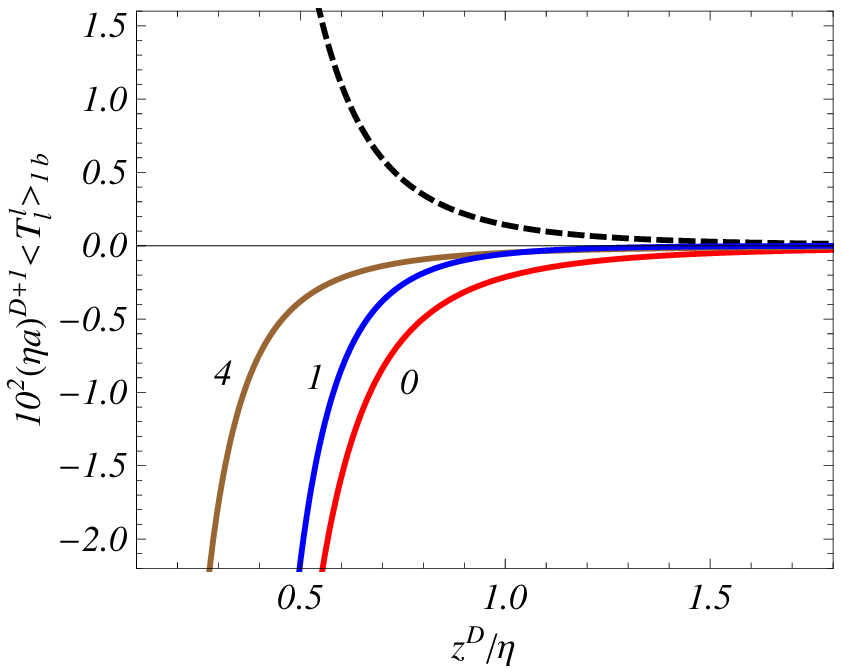,width=7.cm,height=6.cm}%
\end{tabular}%
\end{center}
\caption{Plate-induced parts in the components of the vacuum energy-momentum
tensor as functions of $z^{D}/\protect\eta $ in FRW spacetime with $D=4$.
The full curves correspond to the diagonal components and the dashed one is
for the energy flux. The numbers near the full curves correspond to the
value of the index $l$. For the left plot $c=2/(D+1)$ and for the right one $%
c=2$.}
\label{fig1}
\end{figure}

In figure \ref{fig1} we plot the plate-induced parts in the components of
the vacuum energy-momentum tensor as functions of $z^{D}/\eta $ in FRW
spacetime with $D=4$. The full curves correspond to the diagonal components,
$(\eta a)^{D+1}\langle T_{l}^{l}\rangle _{1b}$, and the dashed one is for
the energy-flux, $(\eta a)^{D+1}\langle T_{0}^{D}\rangle _{1b}$. The numbers
near the full curves correspond to the value of the index $l$. For the left
plot $c=2/(D+1)$ (radiation driven decelerated expansion) and for the right
one we have taken $c=2$ (power law inflation). In the case of the left plot
the parallel stress is negative everywhere. The energy density becomes zero
at $z^{D}/\eta \approx 2.3$. With the further increase of $z^{D}/\eta $, the
energy density takes its maximum value at $\approx 2.85$ and then goes to
zero being positive. This behavior is in accordance with the general
discussion of the asymptotics given before. The normal stress changes the
sign at $z^{D}/\eta \approx 3.7$ and then tends to zero being negative. In
the case of the right plot the energy density and the normal stress do not
change the sign. The parallel stress becomes zero at $z^{D}/\eta \approx 2.1$%
, takes its maximum at $\approx 2.7$ and then goes to zero being positive.

It is also of interest to consider the time-dependence of the plate-induced
parts in the VEVs for an observer with a fixed value of the comoving
coordinate $z^{D}$. The corresponding asymptotic expressions at late and
early stages of the cosmological expansion can be obtained from those given
above for small and large values of $z^{D}/\eta $. For $0<c<1$ and at early
stages of the cosmological expansion, $t\rightarrow 0$, we have $\langle
T_{\mu }^{\mu }\rangle _{1b}\propto t^{2(1-D)c}$ (no sum over $\mu =0,\ldots
,D$), $\langle T_{0}^{D}\rangle _{1b}\propto t^{1-\left( 2D-1\right) c}$. In
this region, for $D>3$ the energy density is positive for $c<3/D$ and
negative for $c>3/D$. At late stages, $t\rightarrow \infty $, the VEVs tend
to zero as: $\langle T_{\mu }^{\mu }\rangle _{1b}\propto t^{-c(D+1)}$ (no
sum over $\mu =0,\ldots ,D-1$), $\langle T_{D}^{D}\rangle _{1b}\propto
t^{-c(D-1)-2}$, $\langle T_{0}^{D}\rangle _{1b}\propto t^{-cD-1}$, and the
corresponding energy density is negative. In the case $c>1$, the behavior of
the VEVs in the limit $t\rightarrow 0$ is as follows: $\langle T_{\mu }^{\mu
}\rangle _{1b}\propto t^{-c(D+1)}$ (no sum over $\mu =0,\ldots ,D-1$), $%
\langle T_{D}^{D}\rangle _{1b}\propto t^{-c(D-1)-2}$ and $\langle
T_{0}^{D}\rangle _{1b}\propto t^{-cD-1}$. At late stages, $t\rightarrow
\infty $, one has: $\langle T_{\mu }^{\mu }\rangle _{1b}\propto t^{-4c}$ (no
sum over $\mu =0,\ldots ,D$), $\langle T_{0}^{D}\rangle _{1b}\propto t^{1-5c}
$. Finally, for models with phantom energy, $c<0$, at early stages of the
expansion, corresponding to $t\rightarrow -\infty $, one gets: $\langle
T_{\mu }^{\mu }\rangle _{1b}\propto (-t)^{-c(D+1)}$ (no sum over $\mu
=0,\ldots ,D-1$), $\langle T_{D}^{D}\rangle _{1b}\propto (-t)^{-c(D-1)-2}$
and $\langle T_{0}^{D}\rangle _{1b}\propto (-t)^{-cD-1}$. In this limit the
energy density goes to $-\infty $, whereas the limiting values of the normal
stress and the energy flux can be either zero or infinity, depending on the
value of the parameter $c$. Near the Big Rip, $t\rightarrow 0$, the
plate-induced parts tend to zero as: $\langle T_{\mu }^{\mu }\rangle
_{1b}\propto (-t)^{-4c}$ (no sum over $\mu =0,\ldots ,D$) and $\langle
T_{0}^{D}\rangle _{1b}\propto (-t)^{1-5c}$. As we see, for all values of $c$%
, the plate-induced parts in the VEV of the energy-momentum tensor decay as
power-law at late stages of the cosmological expansion. At early stages of
the expansion these parts are divergent in models with $c>0$. This is also
the case for the energy density in models with $c<0$. In the latter case the
energy flux at early stages ($t\rightarrow -\infty $) diverges for $c<-1/D$
and vanishes for $c>-1/D$.

\begin{figure}[tbph]
\begin{center}
\epsfig{figure=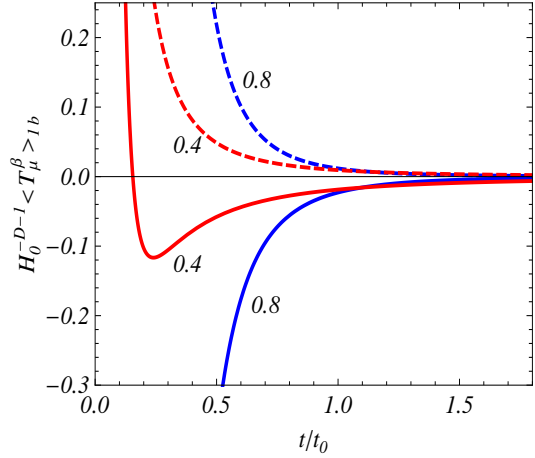,width=7.cm,height=6.cm}
\end{center}
\caption{Plate-induced part in the energy density ($\protect\beta =\protect%
\mu =0$, full curves) and the energy flux ($\protect\beta =D$, $\protect\mu %
=0$, dashed curves) in the model $D=4$, as functions of the comoving time
for a fixed value of comoving coordinate $z^{D}$. The numbers near the
curves correspond to the values of the parameter $c$.}
\label{fig2}
\end{figure}

Figure \ref{fig2} presents the plate-induced part in the energy density ($%
\beta =\mu =0$, full curves) and the energy flux ($\beta =D$, $\mu =0$,
dashed curves) in the model $D=4$, as functions of the comoving time for a
fixed value of comoving coordinate $z^{D}$. The numbers near the curves
correspond to the values of the parameter $c$ and $H_{0}=H(t_{0})$. For the
proper distance from the plate at $t=t_{0}$ we have taken the value $%
a(t_{0})z^{D}=0.5/H_{0}$.

We have considered combined effects of the gravitational field and
boundaries on the properties of the quantum vacuum. Another interesting
physical effect, arising in time-dependent backgrounds, is the creation of
particles. This will give rise to an additional contribution in the
expectation value of the energy-momentum tensor which should be added to the
VEV we have discussed.

\section{The Casimir effect for two conducting plates}

\label{sec:2plates}

For the geometry of two parallel conducting plates we have three separate
regions, $z^{D}<0$, $0<z^{D}<L$, and $z^{D}>L$. In the regions $z^{D}<0$ and
$z^{D}>L$ the VEVs are the same as for a single conducting plate placed at $%
z^{D}=0$ and $z^{D}=L$, respectively. In what follows we consider the region
between the plates (for the electromagnetic Casimir effect for conducting
plates in higher dimensions see, for example, Ref. \cite{Alne07}).

\subsection{Field squared}

As before, first we consider the VEV of the electric field squared. By using
the formula (\ref{FF2pl}) for the two-point function, in the region between
the plates, $0<z^{D}<L$, the VEV is presented in the form

\begin{eqnarray}
\langle E^{2}\rangle &=&\langle E^{2}\rangle _{0}+\frac{8\pi C_{D}}{(\eta
a)^{D+1}}\sum_{n=1}^{\infty }\left[ D-2(1-u_{n})\partial _{u_{n}}\right]
F_{\nu -1}(u_{n})  \notag \\
&&+\frac{4\pi C_{D}}{(\eta a)^{D+1}}\sum_{n=-\infty }^{+\infty }\left[
2(1-y_{n})\partial _{y_{n}}-D+2\right] F_{\nu -1}(y_{n}),  \label{E22pl}
\end{eqnarray}%
with the notations%
\begin{equation}
u_{n}=1-(nL/\eta )^{2},\;y_{n}=1-(z^{D}-nL)^{2}/\eta ^{2}.  \label{un}
\end{equation}%
The $n=0$ term of the last series is the part in the VEV induced by a single
plate at $z^{D}=0$ when the right plate is absent. It coincides with $%
\langle E^{2}\rangle _{1b}$ given by Eq. (\ref{E1pl1b}). Similarly, the $n=1$
term is the VEV induced by a single plate at $z^{D}=L$ when the left plate
is absent. This term is obtained from Eq. (\ref{E1pl1b}) by the replacement $%
z^{D}\rightarrow |z^{D}-L|$. The divergences are contained in these two
terms only and the remaining part is finite for all $0\leqslant
z^{D}\leqslant L$. Consequently, for points near the plates the VEV is
dominated by single plate parts.

A simple expression is obtained for $\nu =1/2$:%
\begin{equation}
\langle E^{2}\rangle =\langle E^{2}\rangle _{0}+\frac{(D-1)\Gamma \left(
(D+1)/2\right) }{(4\pi )^{(D-1)/2}(aL)^{D+1}}\left[ \zeta (D+1)+\frac{3}{2}%
\sum_{n=-\infty }^{+\infty }|z^{D}/L-n|^{-D-1}\right] ,  \label{E2nu12}
\end{equation}%
where $\zeta (x)$ is the Riemann zeta function. The series in this formula
is expressed in terms of the Hurwitz zeta function $\zeta (p,x)$ as:%
\begin{equation}
\sum_{n=-\infty }^{+\infty }|z^{D}/L-n|^{-D-1}=\zeta (D+1,z^{D}/L)+\zeta
(D+1,1-z^{D}/L).  \label{Hurw}
\end{equation}%
Taking $a=1$, we obtain from Eq. (\ref{E2nu12}) the corresponding VEV for
two conducting plates in the Minkowski spacetime. For $D=3$, Eq. (\ref%
{E2nu12}) gives the VEV for general case of the parameter $c$. In both these
cases the plate-induced contribution in the VEV of the electric field
squared is positive everywhere.\qquad

At short separations between the plates, $L/\eta \ll 1$, the argument of the
function $F_{\nu -1}(x)$ is close to 1 and we use the asymptotic formula (%
\ref{FnuAs1}). The leading term of this asymptotic coincides with $%
F_{1/2}(x) $ and, hence, to the leading order $\langle E^{2}\rangle $
coincides with the corresponding VEV in the case $\nu =1/2$, given by (\ref%
{E2nu12}): $\langle E^{2}\rangle \approx \langle E^{2}\rangle _{\nu =1/2}$.
At large separations, $L/\eta \gg 1$, assuming also that $z^{D},L-z^{D}\gg
\eta $, with the help of the asymptotic expression (\ref{Fas2}), for $\nu
\neq 1$ we get%
\begin{eqnarray}
\langle E^{2}\rangle &\approx &\langle E^{2}\rangle _{0}+\frac{2^{2|\nu
-1|}(D-1)\Gamma \left( D/2-|\nu -1|+1\right) }{(4\pi )^{D/2}(\eta
a)^{D+1}(L/\eta )^{D+2-2|\nu -1|}}\Gamma (|\nu -1|)  \notag \\
&&\times \left[ 2\left( |\nu -1|-1\right) \zeta (D+2-2|\nu
-1|)+\sum_{n=-\infty }^{+\infty }\frac{2-|\nu -1|}{|x^{D}/L-n|^{D+2-2|\nu
-1|}}\right] .  \label{E2plAs2}
\end{eqnarray}%
For $\nu =1$ we use Eq. (\ref{F0as2}). The corresponding asymptotic formula
is obtained from Eq. (\ref{E2plAs2}) first replacing $\Gamma (|\nu
-1|)\rightarrow 4\ln (L/\eta )$ and then putting $\nu =0$. In this case the
boundary-induced part of the VEV decays as $\ln (L/\eta )/(L/\eta )^{D+2}$.

For the evaluation of the VEV of the energy-momentum tensor we need also the
VEV of the Lagrangian density. By the calculations similar to those for the
field squared, in the region between two conducting plates we find%
\begin{eqnarray}
\langle F_{\beta \sigma }F^{\beta \sigma }\rangle &=&\langle F_{\beta \sigma
}F^{\beta \sigma }\rangle _{0}+\frac{8\pi C_{D}}{(\eta a)^{D+1}}\left\{
2\sum_{n=1}^{\infty }\left[ 2(1-u_{n})\partial _{u_{n}}-D\right] \left[
F_{\nu -1}(u_{n})-F_{\nu }(u_{n})\right] \right.  \notag \\
&&\left. -\sum_{n=-\infty }^{+\infty }\left\{ \left[ 2(1-y_{n})\partial
_{y_{n}}+D-4\right] F_{\nu }(y_{n})+\left[ 2(1-y_{n})\partial _{y_{n}}-D+2%
\right] F_{\nu -1}(y_{n})\right\} \right\} .  \label{FF2pl1}
\end{eqnarray}%
The $n=0$ and $n=1$ terms of the last series present the single
plate-induced parts for the left and right plates respectively when the
other plate is absent.

\subsection{Energy-momentum tensor}

The VEV of the energy-momentum tensor in the region between the plates is
presented in the form%
\begin{equation}
\langle T_{\mu }^{\mu }\rangle =\langle T_{\mu }^{\mu }\rangle _{0}+\frac{%
C_{D}}{2(\eta a)^{D+1}}\left[ 2\sum_{n=1}^{\infty }g_{\nu }^{(\mu
)}(u_{n})+\sum_{n=-\infty }^{\infty }f_{\nu }^{(\mu )}(y_{n})\right] ,
\label{Tmu2pl}
\end{equation}%
where $f_{\nu }^{(\mu )}(y)$ is defined in Eq. (\ref{fly}) and we have
introduced new functions%
\begin{eqnarray}
g_{\nu }^{(0)}(u) &=&-\left[ 2(1-u)\partial _{u}-D\right] \left[ F_{\nu
}(u)+F_{\nu -1}(u)\right] ,  \notag \\
g_{\nu }^{(l)}(u) &=&\left[ 2\frac{D-3}{D-1}(1-u)\partial _{u}-D+2\right] %
\left[ F_{\nu -1}(u)-F_{\nu }(u)\right] -2F_{\nu }(u),  \label{glu} \\
g_{\nu }^{(D)}(u) &=&\left[ 2(1-u)\partial _{u}-D+2\right] \left[ F_{\nu
}(u)+F_{\nu -1}(u)\right] +2\left( D-3\right) F_{\nu }(u),  \notag
\end{eqnarray}%
with $l=1,\ldots ,D-1$. For the off-diagonal component one has the expression%
\begin{equation}
\langle T_{0}^{D}\rangle =-\frac{2s_{\tau }C_{D}L/\eta }{(\eta a)^{D+1}}%
\sum_{n=-\infty }^{\infty }(z^{D}/L-n)f_{\nu }^{(0D)}(y_{n}),  \label{T0D2pl}
\end{equation}%
with the function $f_{\nu }^{(0D)}(y)$ given by Eq. (\ref{f0D}). The $n=0$
term in the right-hand side of Eq. (\ref{T0D2pl}) gives the energy flux
induced by the plate at $z^{D}=0$ when the right plate is absent ($\langle
T_{0}^{D}\rangle _{1b}$, see Eq. (\ref{T0D1b})). Note that the part induced
by the right plate, $\langle T_{0}^{D}\rangle -\langle T_{0}^{D}\rangle
_{0}-\langle T_{0}^{D}\rangle _{1b}$, vanishes on the surface of the left
plate, at $z^{D}=0$. Similarly, the $n=1$ term of the same series presents
the energy flux in the geometry of a single plate at $z^{D}=L$. From Eq. (%
\ref{T0D2pl}) it follows that the energy flux, considered as a function of $%
z^{D}/L$ is antisymmetric with respect to the point $z^{D}/L=1/2 $ and,
hence, it vanishes at that point.

For $\nu =1/2$ the off-diagonal component vanishes and for the diagonal
components we get simpler expressions (no summation over $\mu $)%
\begin{equation}
\langle T_{\mu }^{\mu }\rangle =\langle T_{\mu }^{\mu }\rangle _{0}-\frac{%
(D-1)\Gamma \left( (D+1)/2\right) }{2(4\pi )^{(D+1)/2}(aL)^{D+1}}\left[
2g_{(0)}^{(\mu )}\zeta (D+1)+\sum_{n=-\infty }^{\infty }\frac{\left(
D-3\right) f_{(0)}^{(\mu )}}{|n-z^{D}/L|^{D+1}}\right] ,  \label{T12}
\end{equation}%
where%
\begin{eqnarray}
g_{(0)}^{(\mu )} &=&1,\;\mu =0,1,\ldots ,D-1,\;g_{(0)}^{(D)}=-D,  \notag \\
f_{(0)}^{(\mu )} &=&1,\;\mu =0,1,\ldots ,D-1,\;f_{(0)}^{(D)}=0.  \label{g0}
\end{eqnarray}%
In particular, for $D=3$ the vacuum energy-momentum tensor is uniform in the
region between the plates. This is a direct consequence of the conformal
invariance of the electromagnetic field in 4-dimensional spacetime.

The expression on the right-hand side of Eq. (\ref{T12}) gives the leading
term in the asymptotic expansion of $\langle T_{\mu }^{\mu }\rangle $ for
small separations between the plates, $L/\eta \ll 1$, for general value of $%
\nu $. In the same limit, for the energy flux one has%
\begin{equation}
\langle T_{0}^{D}\rangle \approx \frac{|1/2-\nu |(D-1)\Gamma ((D+1)/2)}{%
(4\pi )^{(D+1)/2}(a\eta )^{D+1}(L/\eta )^{D}}\sum_{n=-\infty }^{\infty }%
\frac{z^{D}/L-n}{|z^{D}/L-n|^{D+1}}.  \label{T0Dsmall}
\end{equation}%
For large separations between the plates, $L/\eta \gg 1$, and assuming also
that $z^{D},L-z^{D}\gg \eta $, for $\nu <1/2$ we have the asymptotic
expression (no summation over $\mu $)%
\begin{eqnarray}
\langle T_{\mu }^{\mu }\rangle &\approx &\langle T_{\mu }^{\mu }\rangle _{0}+%
\frac{2^{1-2\nu }(D-1)\Gamma (1-\nu )\Gamma \left( D/2+\nu \right) }{(4\pi
)^{D/2+1}(\eta a)^{D+1}(L/\eta )^{D+2\nu }}  \notag \\
&&\times \left[ 2r_{\nu }^{(\mu )}\zeta (D+2\nu )+\sum_{n=-\infty }^{\infty }%
\frac{q_{\nu }^{(\mu )}}{|n-z^{D}/L|^{D+2\nu }}\right] ,  \label{Tmularge}
\end{eqnarray}%
with $q_{\nu }^{(\mu )}$ defined in Eq. (\ref{qunu}) and%
\begin{equation}
r_{\nu }^{(0)}=-\nu ,\;r_{\nu }^{(l)}=\frac{\nu \left( D-3\right) -1}{D-1}%
,\;l=1,\ldots ,D-1,\;r_{\nu }^{(D)}=\nu +1.  \label{rnu}
\end{equation}%
This asymptotic corresponds to the models with $0<c<1$, describing
decelerating expansion.

In the same limit and for $\nu >1/2$ one gets (no sum over $\mu $)%
\begin{eqnarray}
\langle T_{\mu }^{\mu }\rangle &=&\langle T_{\mu }^{\mu }\rangle _{0}+\frac{%
2^{2\nu -1}(D-1)\Gamma (\nu )\Gamma \left( D/2+1-\nu \right) }{(4\pi
)^{D/2+1}(\eta a)^{D+1}(L/\eta )^{D+2-2\nu }}  \notag \\
&&\times \left[ 2s_{\nu }^{(\mu )}\zeta (D+2-2\nu )+\sum_{n=-\infty
}^{\infty }\frac{p_{\nu }^{(\mu )}}{|n-z^{D}/L|^{D+2-2\nu }}\right] ,
\label{Tmularge2}
\end{eqnarray}%
where%
\begin{equation}
s_{\nu }^{(0)}=\nu -1,\;s_{\nu }^{(l)}=\frac{\left( D-3\right) \left( \nu
-2\right) -1}{D-1},\;s_{\nu }^{(D)}=D-\nu -1,  \label{snu}
\end{equation}%
with $l=1,\ldots ,D-1$. This corresponds to accelerating expansion with $c>1$
or $c<-1$.

Now let us consider the asymptotic behavior of the energy flux at large
separations. By using the asymptotic formula (\ref{fnuas}), to the leading
order we find%
\begin{equation}
\langle T_{0}^{D}\rangle =(D-1)\frac{2^{1-2\nu }\Gamma (1-\nu )\Gamma \left(
D/2+1+\nu \right) }{(4\pi )^{D/2+1}(\eta a)^{D+1}(L/\eta )^{D+2\nu }}%
\sum_{n=-\infty }^{\infty }\frac{z^{D}/L-n}{|z^{D}/L-n|^{D+1+2\nu }},
\label{T0Dlarge2pl}
\end{equation}%
for $\nu <1/2$ (this corresponds to $0<c<1$) and
\begin{equation}
\langle T_{0}^{D}\rangle =\frac{2^{2\nu -1}(D-1)\Gamma (\nu )\Gamma \left(
D/2+2-\nu \right) }{(4\pi )^{D/2+1}(\eta a)^{D+1}(L/\eta )^{D+2-2\nu }}%
\sum_{n=-\infty }^{\infty }\frac{z^{D}/L-n}{|z^{D}/L-n|^{D+3-2\nu }},
\label{T0Dlarge2plb}
\end{equation}%
for $\nu >1/2$. The latter case corresponds to $c<0$ or $c>1$ with $s_{\tau
}=-1$.

\begin{figure}[tbph]
\begin{center}
\begin{tabular}{cc}
\epsfig{figure=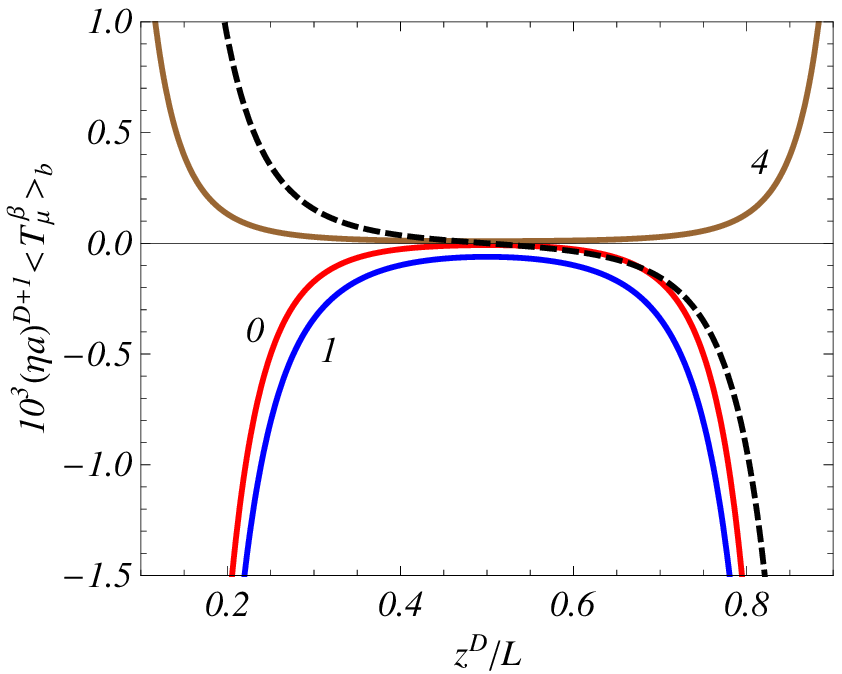,width=7.cm,height=6.cm} & \quad %
\epsfig{figure=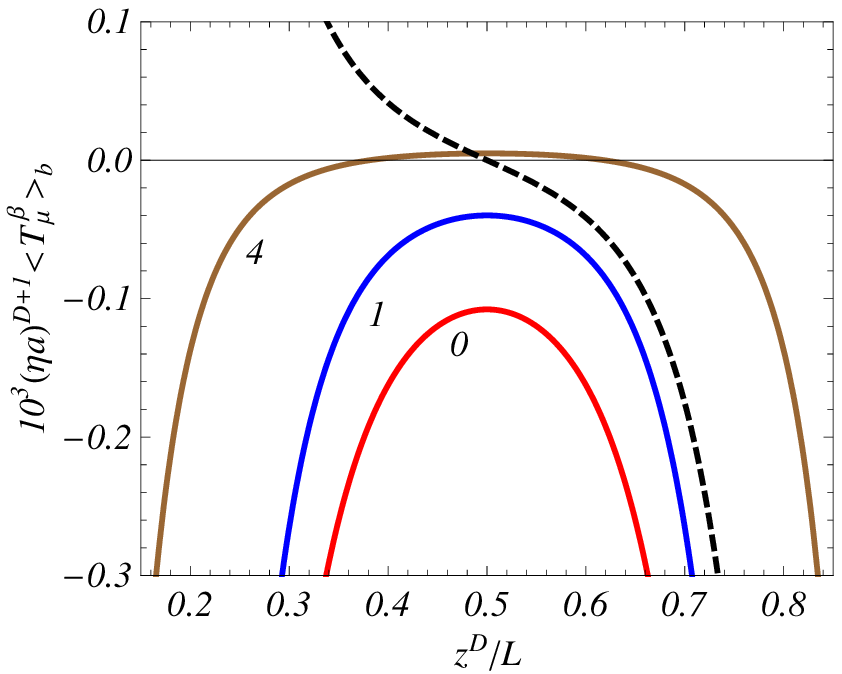,width=7.cm,height=6.cm}%
\end{tabular}%
\end{center}
\caption{Boundary-induced parts in the components of the vacuum
energy-momentum tensor in the region between two conducting plates, as
functions of $z^{D}/\protect\eta $ for a fixed value of $L/\protect\eta =4$.
The graphs are plotted for the FRW model in $D=4$. The full curves
correspond to the diagonal components and the numbers near the curves are
the values of the index $\protect\beta =\protect\mu $. The dashed curves
present the energy flux. For the left panel $c=2/(D+1)$ and for the right
one $c=-3$.}
\label{fig3}
\end{figure}

In figure \ref{fig3} we display the boundary-induced parts in the components
of the vacuum energy-momentum tensor, $\langle T_{\mu }^{\beta }\rangle
_{b}=\langle T_{\mu }^{\beta }\rangle -\langle T_{\mu }^{\beta }\rangle _{0}$%
, in the region between two plates as functions of $z^{D}/\eta $ for a fixed
value of $L/\eta =4$. The graphs are plotted for the FRW model in $D=4$. The
full curves correspond to the diagonal components and the numbers near the
curves are the values of the index $\beta =\mu $. The dashed curve presents
the energy flux. For the left panel $c=2/(D+1)$ and for the right one $c=-3$
(models driven by phantom energy). The graphs for $c=3$ are similar to those
for $c=-3$ with the exception that the normal stress ($\beta =\mu =4$) is
negative everywhere.

\subsection{The Casimir force}

The Casimir force acting on the plate can be obtained evaluating the normal
stress on the surface of the plate. Due to the symmetry of the problem the
forces acting on the left and right plates are equal in magnitude. For the
effective pressure $P$ on the plate at $z^{D}=0$ one has $P=-\langle
T_{D}^{D}\rangle _{z^{D}=0}$. The corresponding force is decomposed into the
self action and interaction parts. The self action part comes from the
stress $\langle T_{D}^{D}\rangle _{1}$. For $D>3$ this part is divergent on
the plate. However, because of the symmetry of the problem, the self-forces
from the left- and right-hand sides of the plate compensate each other and
the corresponding net force vanishes. The interaction part of the vacuum
pressure on the plate is directly obtained from the second term in the
right-hand side of Eq. (\ref{Tmu2pl}) for $\langle T_{D}^{D}\rangle $
omitting the term $n=0$ in the second summation (the latter corresponds to
the self action force). In this way, by taking into account that for $z^{D}=0
$ we have $y_{n}=u_{n}$, for the interaction part of the Casimir pressure
one finds%
\begin{equation}
P^{\mathrm{(int)}}=-\frac{4C_{D}}{(\eta a)^{D+1}}\sum_{n=1}^{\infty }\left\{ %
\left[ (1-u_{n})\partial _{u_{n}}-1\right] F_{\nu }(u_{n})+F_{\nu
-1}(u_{n})/2\right\} .  \label{P}
\end{equation}%
In the special case $\nu =1/2$, this expression is simplified to%
\begin{equation}
P^{\mathrm{(int)}}=-\frac{D(D-1)\Gamma \left( (D+1)/2\right) }{(4\pi
)^{(D+1)/2}(aL)^{D+1}}\zeta (D+1).  \label{P12}
\end{equation}%
This pressure is negative and corresponds to the attractive force. In
particular, for $a=1$ Eq. (\ref{P12}) gives the Casimir force for the plates
in the $(D+1)$-dimensional Minkowski spacetime.

The expression (\ref{P12}) describes the asymptotic behavior of the Casimir
pressure at small separations in general case of $\nu $:%
\begin{equation}
P^{\mathrm{(int)}}\approx P_{\nu =1/2}^{\mathrm{(int)}},\;L/\eta \ll 1.
\label{Psmall}
\end{equation}%
At large separations between the plates, $L/\eta \gg 1$, the dominant
contribution to the Casimir force comes from the term in Eq. (\ref{P}) with
the function $F_{\nu -1}(u_{n})$ for $\nu <1/2$ and from the term with $%
F_{\nu }(u_{n})$ for $\nu >1/2$. By making use of the asymptotic expressions
for these functions, to the leading order we get%
\begin{equation}
P^{\mathrm{(int)}}\approx -\frac{2^{2-2\nu }(D-1)\Gamma (1-\nu )\Gamma
\left( D/2+\nu \right) }{(4\pi )^{D/2+1}(\eta a)^{D+1}(L/\eta )^{D+2\nu }}%
\zeta (D+2\nu ),  \label{Plarge1}
\end{equation}%
for $\nu <1/2$ and
\begin{equation}
P^{\mathrm{(int)}}\approx -\frac{2^{2\nu }(D-1)\left( D-2\nu \right) \Gamma
(\nu )}{(4\pi )^{D/2+1}(\eta a)^{D+1}(L/\eta )^{D+2-2\nu }}\Gamma \left(
D/2+1-\nu \right) \zeta (D+2-2\nu ),  \label{Plarge2}
\end{equation}%
for $\nu >1/2$. In deriving the asymptotics (\ref{Plarge1}) and (\ref%
{Plarge2}) we have assumed that $(L/\eta )^{|2\nu -1|}\gg 1$ and these
asymptotics are not valid for $\nu $ too close to 1/2. Recall that for $\nu
=1/2$ we have the simple expression (\ref{P12}). Note that in all these
cases the effective pressure $P^{\mathrm{(int)}}$ is negative for $\nu <D/2$
and the corresponding vacuum forces are attractive. For $D<2\nu <D+1$ (the
right condition comes from Eq. (\ref{callowed2})), the Casimir pressure
becomes positive and the corresponding forces are repulsive at large
separations. This corresponds to the values of the parameter $c$ in the range%
\begin{equation}
D/3<c<(D-1)/2.  \label{forceneg}
\end{equation}%
Hence, in the range of powers (\ref{forceneg}), the Casimir force is
attractive at small separations and repulsive at large separations. This
means that at some intermediate value for $L$ the force vanishes. This
corresponds to an unstable equilibrium point. Note that, in the range (\ref%
{forceneg}), though the two-point function for the field tensor contains no
infrared divergences, the latter are contained in the two-point function for
the vector potential. In both cases of Eqs. (\ref{Plarge1}) and (\ref%
{Plarge2}), the decay of the Casimir forces with the distance between the
plates is slower than in the case of the Minkowski bulk.

For a fixed value of a comoving separation between the plates, $L$, the
asymptotic behavior of the Casimir force, as a function of the comoving
time, is obtained from Eqs. (\ref{Plarge1}) and (\ref{Plarge2}). For $0<c<1$%
, at early and late stages of the cosmological expansion one has $P^{\mathrm{%
(int)}}\propto t^{-2\left( D-1\right) c}$, $t\rightarrow 0$, and $P^{\mathrm{%
(int)}}\propto t^{-c(D+1)}$, $t\rightarrow \infty $. In the case $c>1$, the
asymptotics are given by $P^{\mathrm{(int)}}\propto t^{-c(D+1)}$ for $%
t\rightarrow 0$ and $P^{\mathrm{(int)}}\propto t^{-4c}$ for $t\rightarrow
\infty $. And finally, in the models driven by phantom energy, $c<0$, the
Casimir force behaves as $P^{\mathrm{(int)}}\propto (-t)^{-c(D+1)}$ for $%
t\rightarrow -\infty $ and $P^{\mathrm{(int)}}\propto (-t)^{-4c}$ for $%
t\rightarrow 0$. In all cases, the force tends to zero at late times and
diverges at early times.

\begin{figure}[tbph]
\begin{center}
\begin{tabular}{cc}
\epsfig{figure=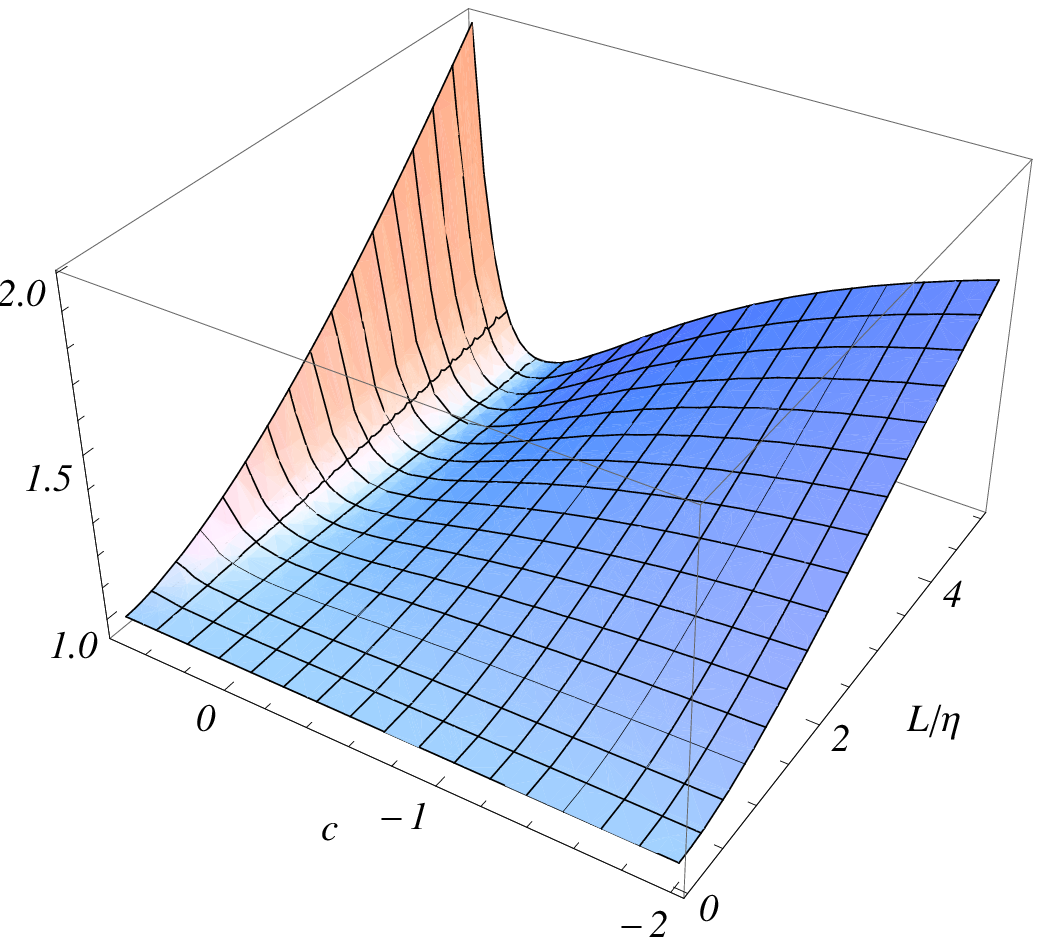,width=7.cm,height=6.cm} & \quad %
\epsfig{figure=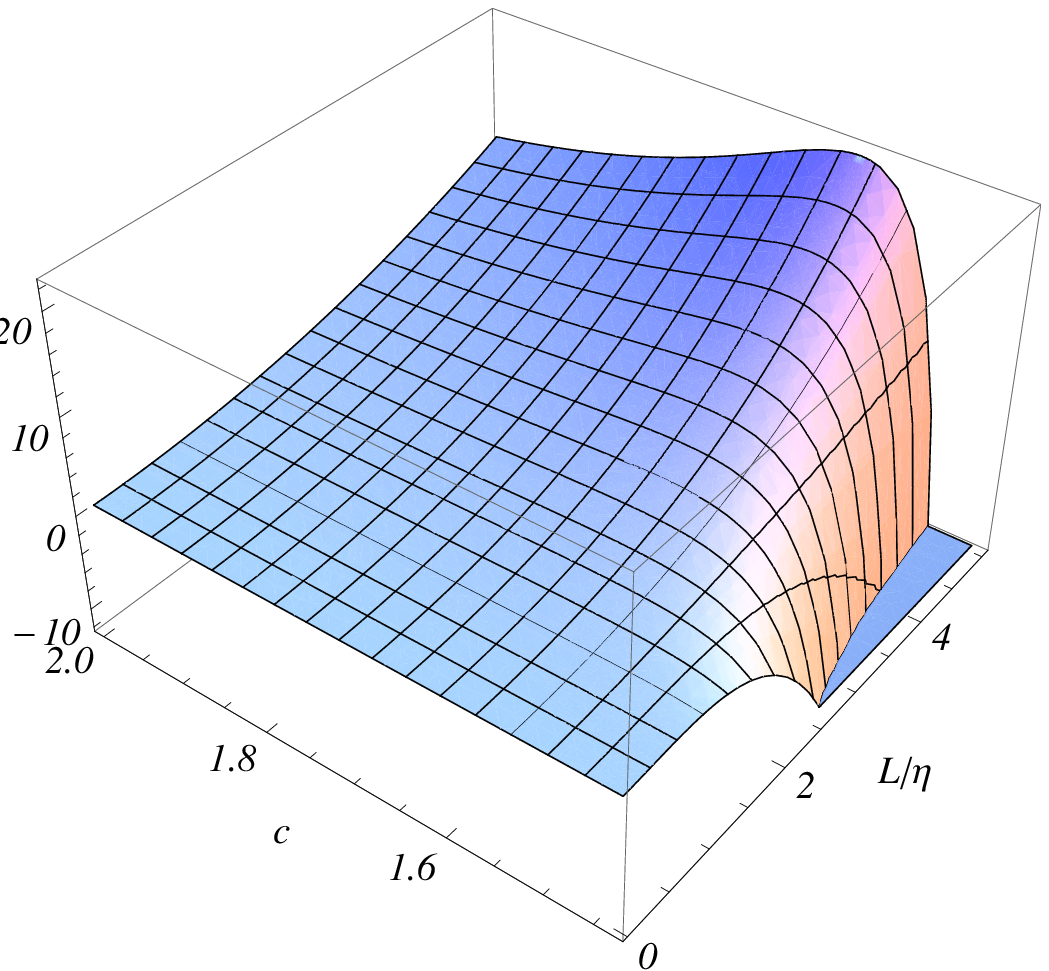,width=7.cm,height=6.cm}%
\end{tabular}%
\end{center}
\caption{Ratio $P^{\mathrm{(int)}}/P_{\protect\nu =1/2}^{\mathrm{(int)}}$ as
a function of $L/\protect\eta $ and of the parameter $c$ in the FRW model
with $D=4$. The left and right plots correspond to the range of powers $%
c<D/(2D-3)$ and $c>D/3$.}
\label{fig4}
\end{figure}

In order to display the influence of the gravitational field on the Casimir
force, in figure \ref{fig4} we present the ratio $P^{\mathrm{(int)}}/P_{\nu
=1/2}^{\mathrm{(int)}}$ as a function of $L/\eta $ and of the parameter $c$
in the FRW model with $D=4$. Note that $P^{\mathrm{(int)}}/P_{\nu =1/2}^{%
\mathrm{(int)}}$ is the ratio of the Casimir pressure in FRW spacetime to
the corresponding pressure for the plates in the Minkowski spacetime with
the separation equal to the proper distance $aL$. The left and right plots
correspond to the range of powers $c<D/(2D-3)$ and $c>D/3$ (see Eq. (\ref%
{callowed2})), respectively. At small separations the effects of the
curvature are small and we have $P^{\mathrm{(int)}}/P_{\nu =1/2}^{\mathrm{%
(int)}}\approx 1$. As it is seen from the graphs, at separations larger than
the curvature radius of the background spacetime the curvature effects are
essential. In particular, in models with power law inflation, depending on
the values of the parameter $c$, the force can change the sign becoming
repulsive.

\section{Conclusion}

\label{sec:Conc}

In this paper we have considered the electromagnetic field two-point
functions in a fixed spatially flat $(D+1)$-dimensional FRW background with
a power law scale factor (\ref{ds2}). As a special case the latter includes
the models of extended inflation ($c>1$) and the models driven by phantom
energy ($c<0$). For the evaluation of the two-point functions the direct
mode summation approach is employed. In this approach a complete set of mode
functions for the vector potential are needed and they are presented in
section \ref{sec:TP}, Eq. (\ref{Amodes}). We assume that the field is
prepared in the Bunch-Davies vacuum state. In the boundary-free geometry,
the corresponding two-point function for the vector potential is given by
Eq. (\ref{AlAm2}). The integral representation of this function is infrared
convergent under the condition (\ref{callowed}) for the parameter $c$.
Though we have no closed expression for the integral in the right-hand side
of Eq. (\ref{AlAm2}), it does not contribute to the two-point function for
the field tensor and, hence, to the VEVs of the field squared and the
energy-momentum tensor. We provided closed expressions, Eq. (\ref{FF0m}),
for the two-point function of the field tensor. This function is infrared
convergent for the values of the parameter $c$ in the range (\ref{callowedF}%
). For the values of $c$ outside this range the Bunch-Davies vacuum in a
boundary-free FRW background is not a physically realizable state.

The two-point functions for the geometry of two parallel perfectly
conducting plates are considered in section \ref{sec:TPpl}. The
corresponding mode functions are given by Eq. (\ref{Amode}) and the
two-point functions are expressed in terms of the boundary-free two-point
functions by Eqs. (\ref{AlAm2pl1}) and (\ref{FF2pl}). The two-point function
of the field tensor for the Bunch-Davies vacuum state is infrared convergent
under the condition (\ref{callowed2}) for the expansion parameter. This
function can be used for the investigation of the VEVs of the field squared
and the energy-momentum tensor. First we have considered the geometry of a
single conducting plate. The corresponding VEVs are decomposed into the
boundary-free and plate-induced contributions. For points outside the plate,
the divergences in the VEVs are the same as in the boundary-free geometry
and the renormalization is reduced to the one for the boundary-free parts in
the VEVs. The plate-induced contributions are finite and they are directly
obtained from the corresponding part in the two-point function for the field
tensor taking the coincidence limit. These contributions depend on the
coordinate $z^{D}$, normal to the plate, in the combination $z^{D}/|\tau |$
with $\tau $ being the conformal time coordinate. Up to a constant factor,
this combination coincides with the ratio of the proper distance from the
plate to the curvature radius of the background spacetime. Simple
expressions for the VEVs are obtained for $\nu =1/2$. This possibility is
realized in two special cases: (i) $c=0$ and $D$ is arbitrary, corresponding
to $(D+1)$-dimensional Minkowski spacetime, and (ii) $D=3$ for general $c$.
In $D=3$ the electromagnetic field is conformally invariant and in this case
the expressions for the Casimir densities in the geometry under
consideration are obtained from those for the Minkowskian bulk by conformal
transformation.

In the geometry of a single plate, the boundary-induced part in the VEV of
the electric field squared is given by Eq. (\ref{E1pl1b}). At small proper
distances from the plate, compared with the curvature radius of the
background spacetime, the effects of the gravitational field are small and
to the leading order the plate-induced part coincides with the corresponding
result for a plate in Minkowski spacetime. In this region the plate-induced
part in the VEV of the electric field squared is positive and it dominates
in the total VEV. At large distances from the plate, the leading terms in
the corresponding asymptotic expansion are given by Eqs. (\ref{E1plAs2}) and
(\ref{E1plAs2b}) for the cases $\nu \neq 1$ and $\nu =1$ respectively.
Depending on the expansion parameter and on the spatial dimension, in this
region the plate-induced part in the VEV of the electric field squared can
be either negative or positive.

For the geometry of a single plate, the boundary-induced contributions in
the VEVs of the diagonal components of the energy-momentum tensor are given
by Eqs. (\ref{Tmu1b}), (\ref{fly}). Note that, unlike to the case of a plate
in the Minkowski spacetime, in the problem under consideration the stresses
along the directions parallel to the plate are not equal to the energy
density. In addition to the diagonal components, the vacuum energy-momentum
tensor has also a nonzero off-diagonal component which corresponds to the
energy flux along the direction normal to the plate. The boundary-free part
in the VEV of the energy-momentum tensor is diagonal and the energy flux is
purely boundary-induced effect. For points near the plate, the leading terms
in the asymptotic expansion of the energy-momentum tensor components are
given by Eqs. (\ref{TDDas1}) and (\ref{T0Das1}). In this region the energy
density and the parallel stresses are negative and the off-diagonal
component $\langle T_{0}^{D}\rangle _{1b}$ is positive. The latter means
that the energy flux is directed from the plate. The normal stress is
positive in models with decelerating expansion and negative for accelerating
expansion. At large proper distances from the plate compared with the
curvature radius of the background spacetime, the leading terms in the
plate-induced VEVs are given by Eqs. (\ref{Tlarge1}), (\ref{T0Dlarge1}) in
the case of decelerated expansion and by Eqs. (\ref{Tlarge2}), (\ref%
{T0Dlarge}) in the case of accelerated expansion. In both cases, the decay
of the plate-induced parts in the energy density and in the parallel
stresses, as functions of the distance from the plate, is slower than in the
corresponding problem on Minkowski bulk. Depending on the value of the
expansion parameter, the plate-induced contribution in the energy density
can be either negative or positive. The energy flux is always directed from
the plate. For a fixed value of the comoving coordinate $z^{D}$, at late
times of the cosmological expansion, the plate-induced parts in the VEV of
the energy-momentum tensor decay as power-law. In particular, the energy
density decays as $t^{-(D+1)c}$ in models with decelerated expansion ($0<c<1$%
) and like $|t|^{-4c}$ for an accelerated expansion.

In the region between two parallel conducting plates the VEVs are decomposed
as Eqs. (\ref{E22pl}), (\ref{Tmu2pl}), and (\ref{T0D2pl}) for the field
squared and the energy-momentum tensor, respectively. The $n=0$ and $n=1$
terms in the last series of these representations correspond to the single
plate parts (left and right, respectively) when the second plate is absent.
The surface divergences are contained in these terms only and the remained
part is finite on the plates. Single plate parts dominate near the plates.
The energy flux, considered as a function of $z^{D}$, is antisymmetric with
respect to the point $z^{D}=L/2$. It is directed from the left plate for $%
0<z^{D}<L/2$ and from the right plate for $L/2<z^{D}<L$ and vanishes at $%
z^{D}=L/2$. We have provided simple asymptotic expressions of the VEVs for
small and large separations between the plates. For the geometry of a single
plate the Casimir normal stresses on the left- and right-hand sides are the
same and the net force acting on the plate vanishes. The presence of the
second plate induces a force referred here as an interaction force. The
latter, acting on the left plate, is obtained putting $z^{D}=0$ in the part
of the normal stress induced by the right plate. The corresponding effective
pressure is given by Eq. (\ref{P}). At small separations between the plates,
the spacetime curvature effects are subdominant and to the leading order we
recover the result for the plates in the Minkowski spacetime, Eq. (\ref{P12}%
). In this limit the forces are attractive. At larger separations, the
asymptotics of the Casimir forces are given by Eqs. (\ref{Plarge1}) and (\ref%
{Plarge2}) for decelerated and accelerated expansions respectively. The
corresponding forces are attractive except the range of powers (\ref%
{forceneg}), where the Casimir pressure becomes positive at large
separations and the corresponding forces are repulsive. In this range for
the values of the parameter $c$, though the two-point function of the field
tensor contains no infrared divergences, the latter are contained in the
two-point function for the vector potential.

\section*{Acknowledgments}

This work was supported by State Committee Science MES RA, within the frame
of the research project No. SCS 13-1C040. S. B. was partly supported by the
ERC Advanced Grant No. 226455, "Supersymmetry, Quantum Gravity and Gauge
Fields" (SUPERFIELDS). A. A. S. gratefully acknowledges the hospitality of
the INFN, Laboratori Nazionali di Frascati (Frascati, Italy), where part of
this work was done.

\end{document}